\documentclass{aastex631}

\usepackage{amsthm,amsmath,amssymb}
\usepackage{lipsum}
\usepackage{float}
\usepackage{soul}

\begin{document}

\title{BSN-II: The First Light Curve Study of Eight Total Eclipsing Contact Binary Stars with Shallow Fillout Factors}

\author[0000-0002-0196-9732]{Atila Poro}
\altaffiliation{atilaporo@bsnp.info, atila.poro@obspm.fr (AP)}
\affiliation{Binary Systems of South and North (BSN) Project, 15875 Tehran, Iran}
\affiliation{LUX, Observatoire de Paris, CNRS, PSL, 61 Avenue de l'Observatoire, 75014 Paris, France}
\affiliation{Astronomy Department of the Raderon AI Lab., BC., Burnaby, Canada}

\author[0000-0003-3590-335X]{Kai Li}
\affiliation{School of Space Science and Technology, Institute of Space Sciences, Shandong University, Weihai, Shandong 264209, People's Republic of China}

\author[0000-0003-1263-808X]{Raul Michel}
\affiliation{Instituto de Astronom\'ia, UNAM. A.P. 106, 22800 Ensenada, BC, M\'exico}

\author[0009-0005-0485-418X]{Li-Heng Wang}
\affiliation{School of Space Science and Technology, Institute of Space Sciences, Shandong University, Weihai, Shandong 264209, People's Republic of China}

\author[0000-0002-1972-8400]{Fahri Alicavus}
\affil{Çanakkale Onsekiz Mart University, Faculty of Sciences, Department of Physics, 17020, Çanakkale, Türkiye}
\affil{Çanakkale Onsekiz Mart University, Astrophysics Research Center and Ulupnar Observatory, 17020, Çanakkale,
Türkiye}

\author{Morgan Rhai Nájera}
\affil{Instituto de Astronom\'ia, UNAM. A.P. 106, 22800 Ensenada, BC, M\'exico}

\author{Priscila Santill\'an-Ortega}
\affiliation{Instituto de Astronom\'ia, Universidad Nacional Aut\'onoma de M\'exico, Ciudad Universitaria 04510, CDMX, M\'exico}

\author[0000-0002-9761-9509]{Francisco Javier Tamayo}
\affiliation{Facultad de Ciencias F\'{\i}sico-Matem\'aticas, UANL, 66451 San Nicol\'as de los Garza, NL, M\'exico}

\author[0000-0002-7348-8815]{Hector Aceves}
\affiliation{Instituto de Astronom\'ia, UNAM. A.P. 106, 22800 Ensenada, BC, M\'exico}

\begin{abstract}
This study provides the first comprehensive analysis of eight total-eclipse W Ursae Majoris-type contact binary systems. Ground-based photometric multiband observations were conducted at a Mexican observatory, and new times of minima were extracted. The O-C analysis reveals that four of our target binaries exhibit a long-term increase in their orbital periods, while the others show a long-term decrease in their orbital periods. We analyzed the light curves using the PHOEBE Python code and BSN application. Among the target systems, two required the inclusion of a cold starspot on one of the components to achieve an adequate fit. The light curve analysis revealed that the target systems exhibit a shallow fillout factor. Absolute parameters were estimated using the Gaia DR3 parallax and astrophysics equations. Considering the effective temperatures and component masses, each system was classified as either the A- or W-subtype. The stellar evolution of the systems was represented through the mass-radius and mass-luminosity diagrams. Additionally, we calculated the initial masses of the companion stars and the total mass lost for each target system.
\end{abstract}


\section{Introduction}
\label{sec1}
The W Ursae Majoris contact binary systems are composed of two late-type stars that share a common convective envelope, enabling continuous mass and energy exchange (\citealt{1968ApJ...151.1123L}, \citealt{1968ApJ...153..877L}). These systems display light curves with nearly equal minima and smooth, continuous brightness variations, which are indicative of their overcontact configuration and the persistent mutual eclipses between the components (\citealt{kuiper1941}, \citealt{2014ApJS..212....4Q}).

The orbital period distribution of contact binaries peaks near 0.27 days, with a well-established short-period cutoff around 0.22 days, possibly reflecting a limit imposed by angular momentum loss or structural instability in low-mass components (\citealt{2012MNRAS.421.2769J}). The period evolution is closely related to both mass transfer and angular momentum loss mechanisms such as magnetic braking (\citealt{2013ApJS..209...13Q}).

The mass ratio ($q=M_2/M_1$) is a fundamental parameter in characterizing contact binary systems, as it significantly affects the system's stability, light curve morphology, and mass transfer dynamics. However, determining the mass ratio remains challenging, particularly in systems where one component has a very small mass and radius (\citealt{2022AJ....164..202L}).

A notable photometric feature observed in many contact binary systems is the O'Connell effect (\citealt{1951PRCO....2...85O}), which manifests as unequal maxima in the light curve outside of eclipse. While its origin is still debated, proposed mechanisms include the presence of cold or hot spots on the stellar surfaces, likely associated with magnetic activity (\citealt{2003ChJAA...3..142L}).

Another critical parameter is the degree of contact, often quantified by the fillout factor ($f$), which measures the extent to which the stars overfill their Roche lobes. The fillout factor \( f \) is defined as

\begin{equation}
\begin{aligned}
f = \frac{\Omega - \Omega_{\mathrm{in}}}{\Omega_{\mathrm{out}} - \Omega_{\mathrm{in}}},
\end{aligned}
\end{equation}

\noindent where \( \Omega \) is the surface potential of the common envelope, and \( \Omega_{\mathrm{in}} \), \( \Omega_{\mathrm{out}} \) are the inner and outer critical potentials, respectively (\citealt{mochnacki1981}).

Contact binary systems are conventionally classified into two subtypes: A-type and W-type. In A-type systems, the more massive component exhibits a higher effective temperature, whereas in W-type systems, the more massive component is comparatively cooler (\citealt{1970VA.....12..217B}). Subsequent analysis by \cite{2020MNRAS.492.4112Z} indicated that these subtypes evolve through distinct evolutionary pathways. The classification not only reflects structural and thermal distinctions, but also correlates with angular momentum loss and mass transfer behavior within the systems \citep{1970VA.....12..217B, 2020RAA....20..163Q}.

This study employed ground-based multiband photometric observations of eight totally eclipsing W UMa-type contact binaries to obtain more accurate orbital and physical parameters. This work continues the investigation initiated by \cite{2025MNRAS.537.3160P}, providing additional observations and analysis of the characteristics of other W UMa-type contact binaries in the BSN\footnote{\url{https://bsnp.info/}} project. The structure of this paper is as follows: Section 2 provides information about the target systems. Section 3 describes the ground-based observations and data reduction process. Orbital period variations are discussed in Section 4. Section 5 presents the photometric light curve solutions for the target systems. The methods and outcome used to determine the absolute parameters are detailed in Section 6. Finally, discussion and conclusions are presented in Section 7.

\vspace{0.6cm}
\section{Target Systems}
\label{sec2}
We have analyzed eight eclipsing binary stars, including CRTS J165528.6+294254 (hereinafter J1655), CRTS J170839.8+122530 (hereinafter J1708), CRTS J170956.7+164054 (hereinafter J1709), CRTS J224931.9+314743 (hereinafter J2249), LINEAR 20334947 (hereinafter L2033), NSVS 8849526 (hereinafter N8849), WISE J200342.3+363643 (hereinafter W2003), and WISE J201926.1+461759 (hereinafter W2019). Table \ref{systemsinfo} provides specifications for the target systems from the Gaia DR3 database (\citealt{2023AA...674A..33G}), along with system temperatures obtained from the TESS Input Catalog (TIC) v8.2 database. The general properties of the target systems are as follows:

$\bullet$ J1655: This binary system was discovered in the Catalina Surveys Data Release 1 (CSDR1; \citealt{2014ApJS..213....9D}). The Zwicky Transient Facility (ZTF, \citealt{2023AA...675A.195S}) catalog of periodic variable stars provided an ephemeris for this system: $2458236.446519^{\mathrm{HJD}}+0.2699958 \times E$. Additionally, the All-Sky Automated Survey for Supernovae (ASAS-SN; \citealt{2018MNRAS.477.3145J}) reported an orbital period of 0.2699918 days. The ASAS and Variable Star Index (VSX) catalogs listed a maximum magnitude of $16.398^{mag}$ for this binary system.

$\bullet$ J1708: This system was discovered as a contact binary in the CSDR1 catalog (\citealt{2014ApJS..213....9D}). The CSDR1, VSX, ZTF, and ASAS-SN catalogs consistently classified J1708 as a contact binary with an orbital period of 0.28299 days. The VSX database reported a maximum magnitude of $14.790^{mag}$ for J1708.

$\bullet$ J1709: CSDR1 (\citealt{2014ApJS..213....9D}) discovered this binary system and ZTF catalog (\citealt{2020ApJS..249...18C}) listed it as a contact binary system. The CSDR1, VSX, ZTF, and ASAS-SN catalogs reported an orbital period of 0.27775 days. The VSX database lists a maximum magnitude of $16.030^{mag}$ with an amplitude of $0.620^{mag}$. The effective temperature of this system is not reported in Gaia DR3; however, Gaia DR2 reported a temperature of 4980 K, which differs by 259 K from the value provided by TIC (Table \ref{systemsinfo}).

$\bullet$ J2249: This system was discovered by the CSDR1 catalog (\citealt{2014ApJS..213....9D}). J2249 is classified as a contact binary system in the CSDR1, ZTF, ASAS-SN, and Asteroid Terrestrial-impact Last Alert System (ATLAS) catalogs. The orbital period of this system is listed as 0.31506 days in the ASAS-SN, ZTF, and VSX catalogs. Additionally, the VSX catalog provides a maximum magnitude of $14.780^{mag}$ for J2249. The effective temperature of this system is not available in Gaia DR3; however, Gaia DR2 lists a temperature of 4950 K, which is close to the value provided by TIC.

$\bullet$ L2033: This system is located in the Hercules constellation in the northern hemisphere. The L2033 binary system was discovered by the Lincoln Near-Earth Asteroid Research (LINEAR; \citealt{2013AJ....146..101P}). The VSX database reported an orbital period of 0.278594 days and a maximum magnitude of $16.080^{mag}$ for L2033.

$\bullet$ N8849: This binary system was discovered by the Robotic Optical Transient Search Experiment I (ROTSE-I) telescope, which presented the first results of a search for periodic variable stars (\citealt{2000AJ....119.1901A}, \citealt{2006AJ....131..621G}). N8849 was reported in the Northern Sky Variability Survey (NSVS; \citealt{2009AJ....138..466H}) as a contact binary system. This system, located in the Pegasus constellation, is reported in the ASAS-SN, ZTF, and VSX catalogs as a contact binary, with an orbital period of 0.28572 days and a maximum apparent magnitude of $13.164^{mag}$, as listed in the VSX database.

$\bullet$ W2003: This system was classified as a contact binary system by the Wide-field Infrared Survey Explorer (WISE, \citealt{2018ApJS..237...28C}) catalog. The maximum apparent magnitude of this system is $14.651^{mag}$, and its orbital period is 0.2929868 days, as reported in the VSX database. The 355 K temperature difference between the values reported by Gaia DR3 and TIC is considerable for W2003.

$\bullet$ W2019: This system is also known as ZTF J201926.16+461758.9 and is classified as a contact binary in the WISE catalog (\citealt{2018ApJS..237...28C}). The orbital period and maximum apparent magnitude of this system in the VSX database are 0.2951102 days and $14.493^{mag}$, respectively.

\begin{table*}
\renewcommand\arraystretch{1.2}
\caption{Specifications of the target systems from the Gaia DR3 and the effective temperature of TIC.}
\centering
\begin{center}
\footnotesize
\begin{tabular}{c c c c c c c c}
\hline
System & RA$.^\circ$(J2000) & Dec$.^\circ$(J2000) & $d$(pc) & RUWE & $V-R$(mag) & $T_{Gaia}$(K) & $T_{TIC}$(K)\\
\hline
CRTS J165528.6+294254 (J1655)	&	253.869459	&	29.715025	&	938(24)	&	1.034	& 0.557 & 4817(8)	&	4737(165)	\\
CRTS J170839.8+122530 (J1708)	&	257.165979	&	12.425025	&	780(14)	&	1.047	& 0.473 & 5100(7)	&	5024(182)	\\
CRTS J170956.7+164054 (J1709)	&	257.486704	&	16.682181	&	1436(74)	&	1.011	& 0.489 &	-	&	5239(200)	\\
CRTS J224931.9+314743 (J2249)	&	342.383260	&	31.795409	&	831(16)	&	0.985	& 0.521 &	-	&	4983(168)	\\
LINEAR 20334947 (L2033)	&	259.936894	&	32.555602	&	1182(44)	&	0.966	& 0.576 &	4804(35)	&	4668(153)	\\
NSVS 8849526 (N8849)	&	330.499640	&	33.613291	&	397(2)	&	1.300	& 0.519 &	5257(34)	&	5253(178)	\\
WISE J200342.3+363643 (W2003)	&	300.926389	&	36.612068	&	729(10)	&	1.079	& 0.635 &	5656(27)	&	5301(196)	\\
WISE J201926.1+461759 (W2019)	&	304.858989	&	46.299672	&	816(11)	&	1.080	& 0.429 &	-	&	5321(200)	\\
\hline
\end{tabular}
\end{center}
\label{systemsinfo}
\end{table*}

\vspace{0.6cm}
\section{Observation and Data Reduction}
\label{sec3}
Ground-based observations of the eight binary systems were carried out at the San Pedro Mártir (SPM) Observatory in México, located at $115^\circ$ $27^{'}$ $49^{''}$ West, $31^\circ$ $02^{'}$ $39^{''}$ North, at an altitude of 2830 meters above sea level. The observations were conducted using a 0.84-meter Ritchey-Chrétien telescope with an $f/15$ focal ratio, equipped with a Marconi-5 CCD detector from Spectral Instruments featuring an e2v CCD231-42 chip with $15\times15\mu m2$ pixels, a gain of $2.2 e-\diagup$ ADU, and a readout noise of $3.6 e-$. Observations were performed using $B$, $V$, $R_c$, and $I_c$ standard filters. Photometric images were processed using IRAF routines, as described by \cite{1986SPIE..627..733T}. Data reduction included bias subtraction and flat-field correction.

Table \ref{observations} presents the observational details for each target system, including the observation date, filters used, and exposure times. Table \ref{stars} provides the general characteristics of the comparison and check stars used during the observation and data reduction processes. The information in Table \ref{stars} is from Gaia DR3 (\citealt{2023AA...674A..33G}).

\begin{table*}
\renewcommand\arraystretch{1.2}
\caption{Specifications of the ground-based observations.}
\centering
\begin{center}
\footnotesize
\begin{tabular}{c c c c}
\hline
System & Observation Date & Filter & Exposure time(s)\\
\hline
J1655	&	 2024 (May 21) 	&	 $VR_cI_c$ 	&	 $V(100)$, $R_c(60)$, $I_c(50)$ 	\\
J1655	&	 2024 (June 1) 	&	 $BVR_cI_c$ 	&	 $B(70)$, $V(50)$, $R_c(35)$, $I_c(30)$ 	\\
J1708	&	 2024 (July 1) 	&	 $BVR_cI_c$ 	&	 $B(60)$, $V(40)$, $R_c(25)$, $I_c(20)$ 	\\
J1709	&	 2024 (June 30) 	&	 $VR_cI_c$ 	&	 $V(90)$, $R_c(50)$, $I_c(40)$ 	\\
J2249	&	 2024 (September 6) 	&	 $BVR_cI_c$ 	&	 $B(90)$, $V(50)$, $R_c(35)$, $I_c(30)$ 	\\
L2033	&	 2024 (June 28) 	&	 $VR_cI_c$ 	&	 $V(90)$, $R_c(50)$, $I_c(40)$ 	\\
N8849	&	 2024 (August 17) 	&	 $BVR_cI_c$ 	&	 $B(90)$, $V(50)$, $R_c(35)$, $I_c(30)$ 	\\
W2003	&	 2024 (July 22, July 30) 	&	 $BVR_cI_c$ 	&	 $B(90)$, $V(50)$, $R_c(35)$, $I_c(30)$ 	\\
W2019	&	 2024 (July 24) 	&	 $BVR_cI_c$ 	&	 $B(90)$, $V(50)$, $R_c(35)$, $I_c(30)$ 	\\
\hline
\end{tabular}
\end{center}
\label{observations}
\end{table*}

\begin{table*}
\renewcommand\arraystretch{1.2}
\caption{List the comparisons and check stars in the ground-based observations.}
\centering
\begin{center}
\footnotesize
\begin{tabular}{c c c c c}
\hline
System & Star type & RA$.^\circ$(J2000) & DEC$.^\circ$(J2000) & $V-R$(mag)\\
\hline
J1655	&	Comparison	&	253.905801	&	29.760224	&	0.593	\\
J1655	&	Check	&	253.826437	&	29.680841	&	0.386	\\
J1708	&	Comparison	&	257.210742	&	12.373291	&	0.529	\\
J1708	&	Check	&	257.222920	&	12.466926	&	0.526	\\
J1709	&	Comparison	&	257.475647	&	16.691172	&	0.505	\\
J1709	&	Check	&	257.477029	&	16.703282	&	0.555	\\
J2249	&	Comparison	&	342.397622	&	31.763124	&	0.426	\\
J2249	&	Check	&	342.332032	&	31.828449	&	0.376	\\
L2033	&	Comparison	&	259.984549	&	32.644869	&	0.565	\\
L2033	&	Check	&	260.013320	&	32.527909	&	0.510	\\
N8849	&	Comparison	&	335.411845	&	28.090665	&	0.500	\\
N8849	&	Check	&	335.507518	&	28.104244	&	0.510	\\
W2003	&	Comparison	&	300.978121	&	36.619803	&	0.602	\\
W2003	&	Check	&	300.892047	&	36.604192	&	0.600	\\
W2019	&	Comparison	&	304.863032	&	46.311388	&	0.491	\\
W2019	&	Check	&	304.874146	&	46.292615	&	0.456	\\
\hline
\end{tabular}
\end{center}
\label{stars}
\end{table*}

\vspace{0.6cm}
\section{Orbital Period Variations}
\label{sec4}
\begin{figure*}
\centering
\includegraphics[width=0.47\textwidth]{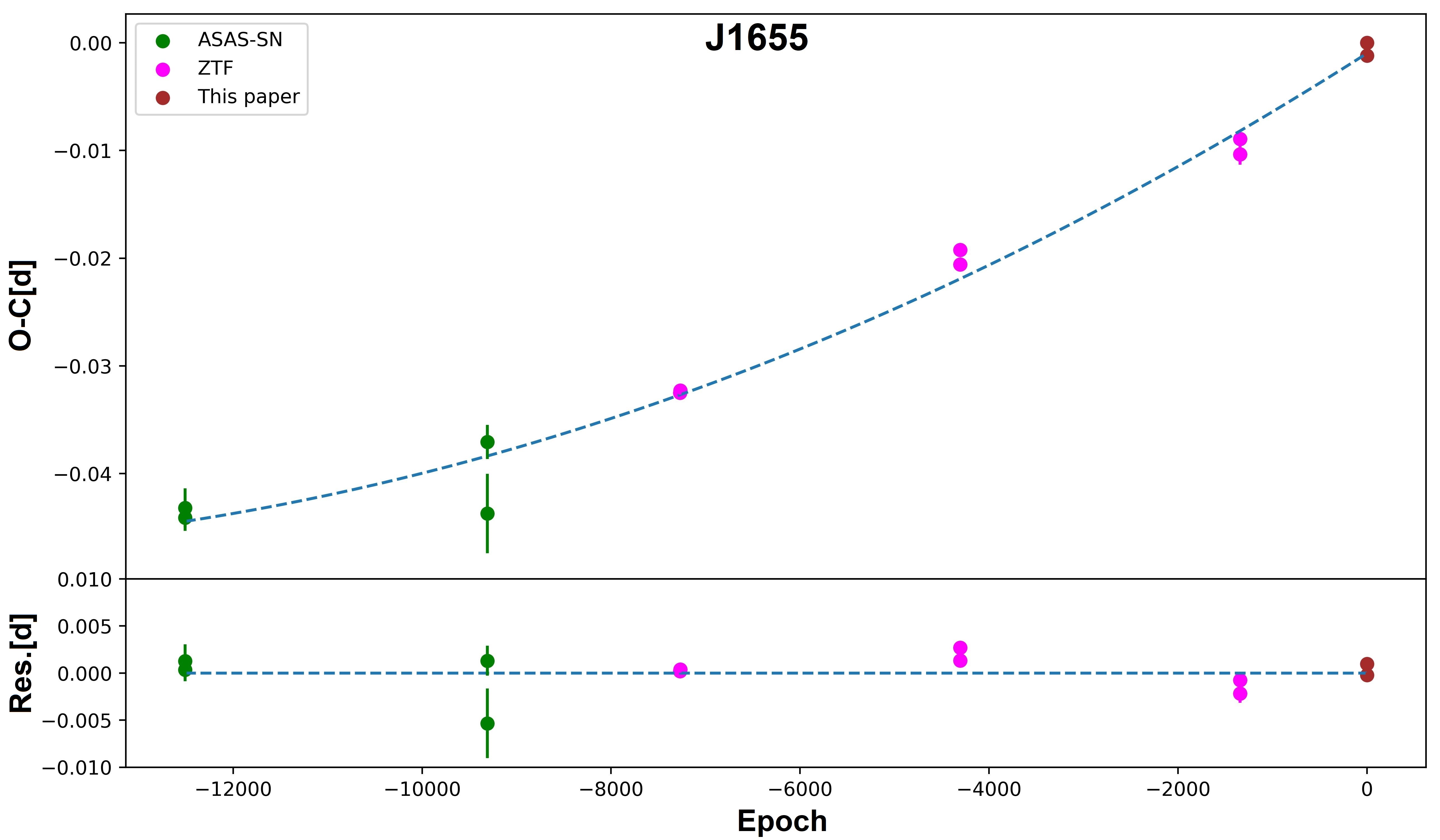}
\includegraphics[width=0.47\textwidth]{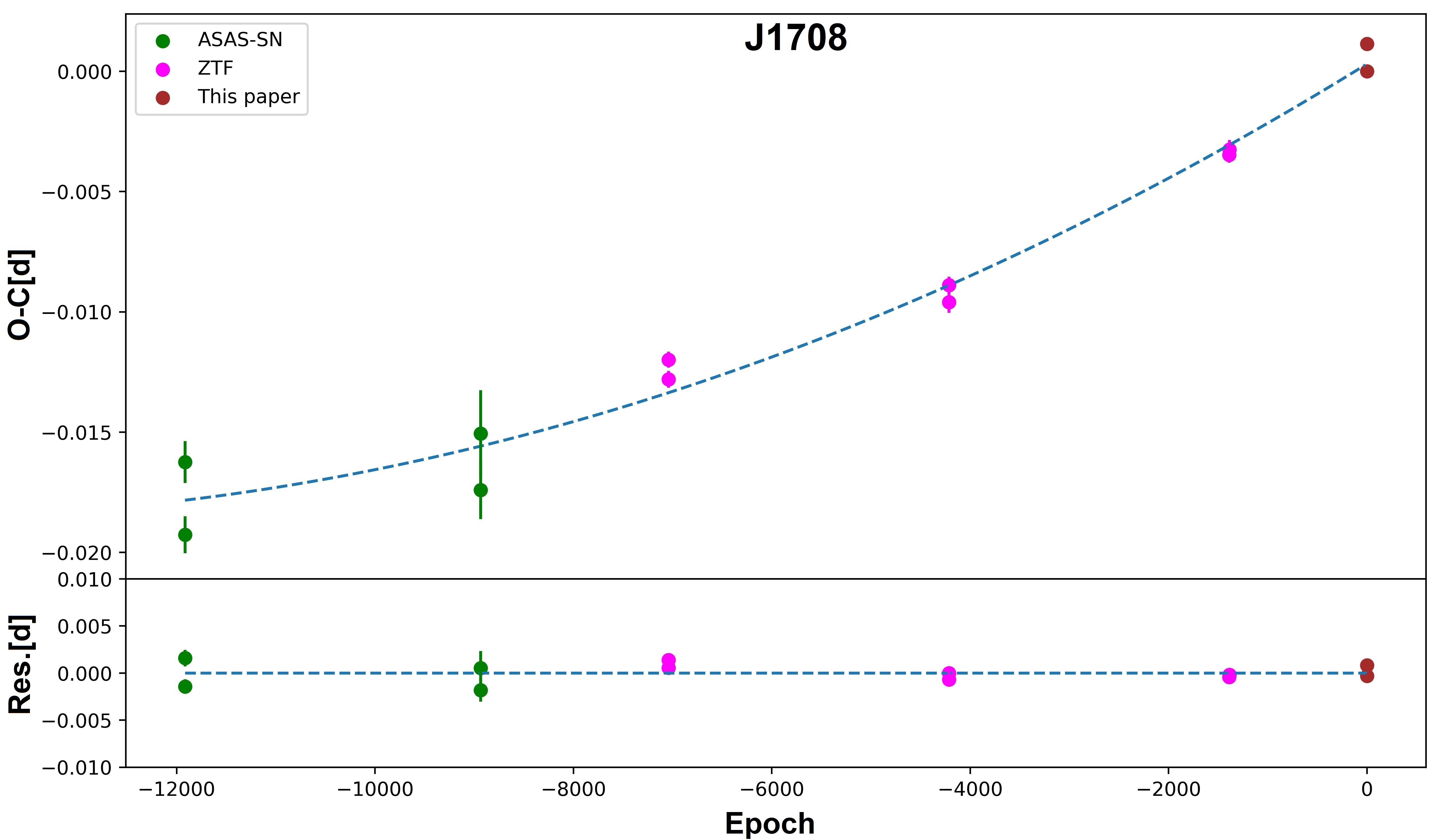}
\includegraphics[width=0.47\textwidth]{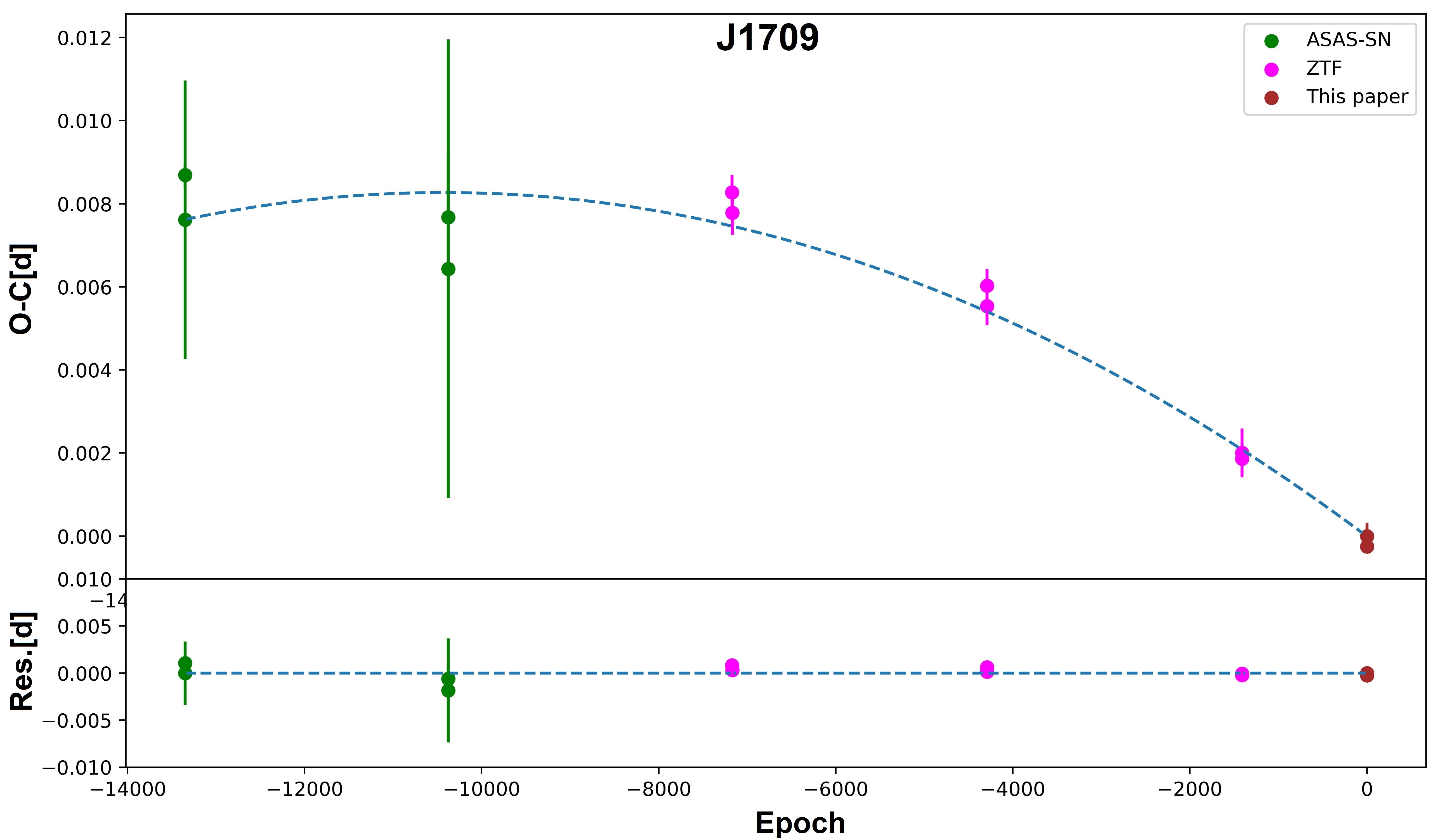}
\includegraphics[width=0.47\textwidth]{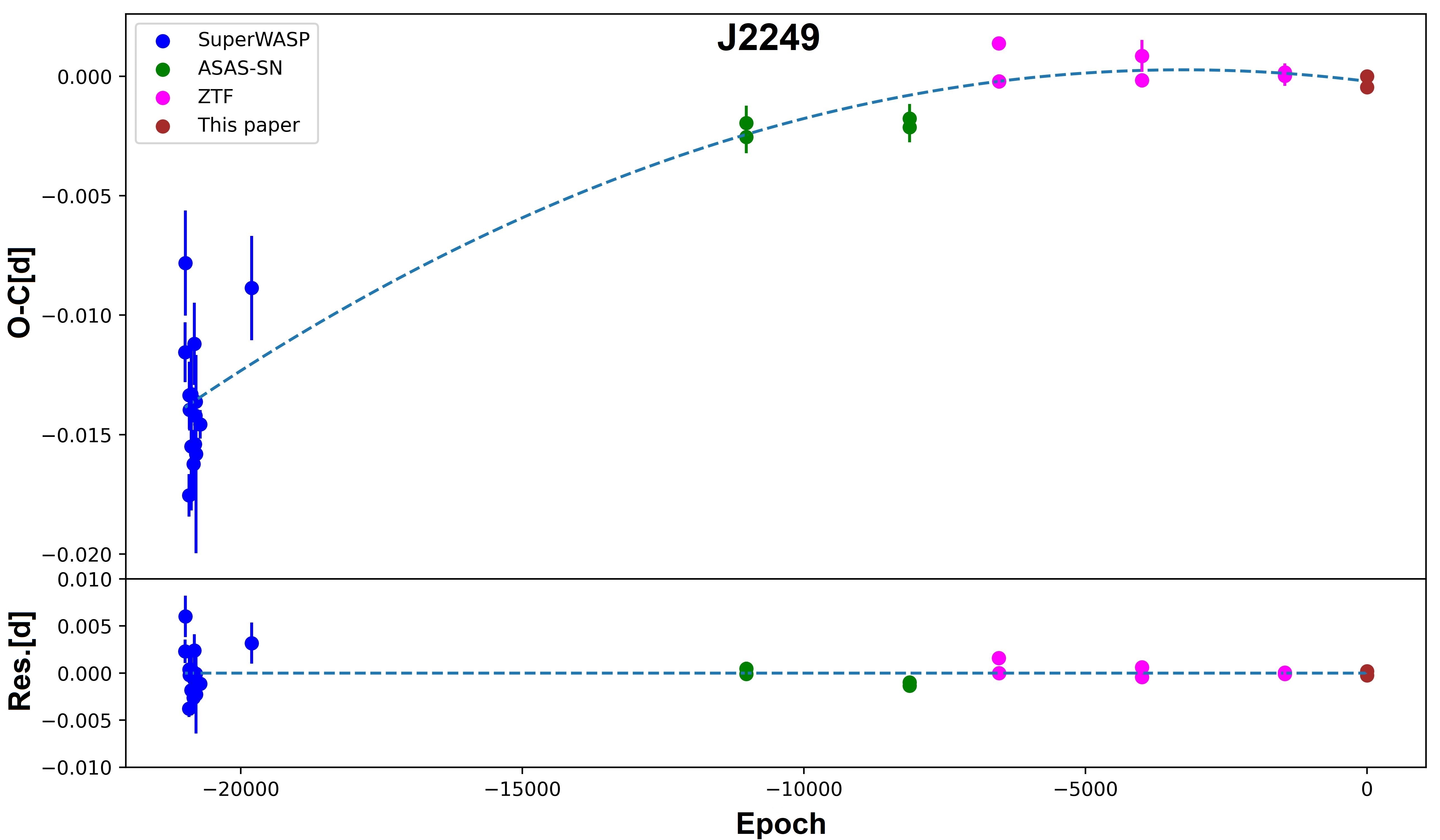}
\includegraphics[width=0.47\textwidth]{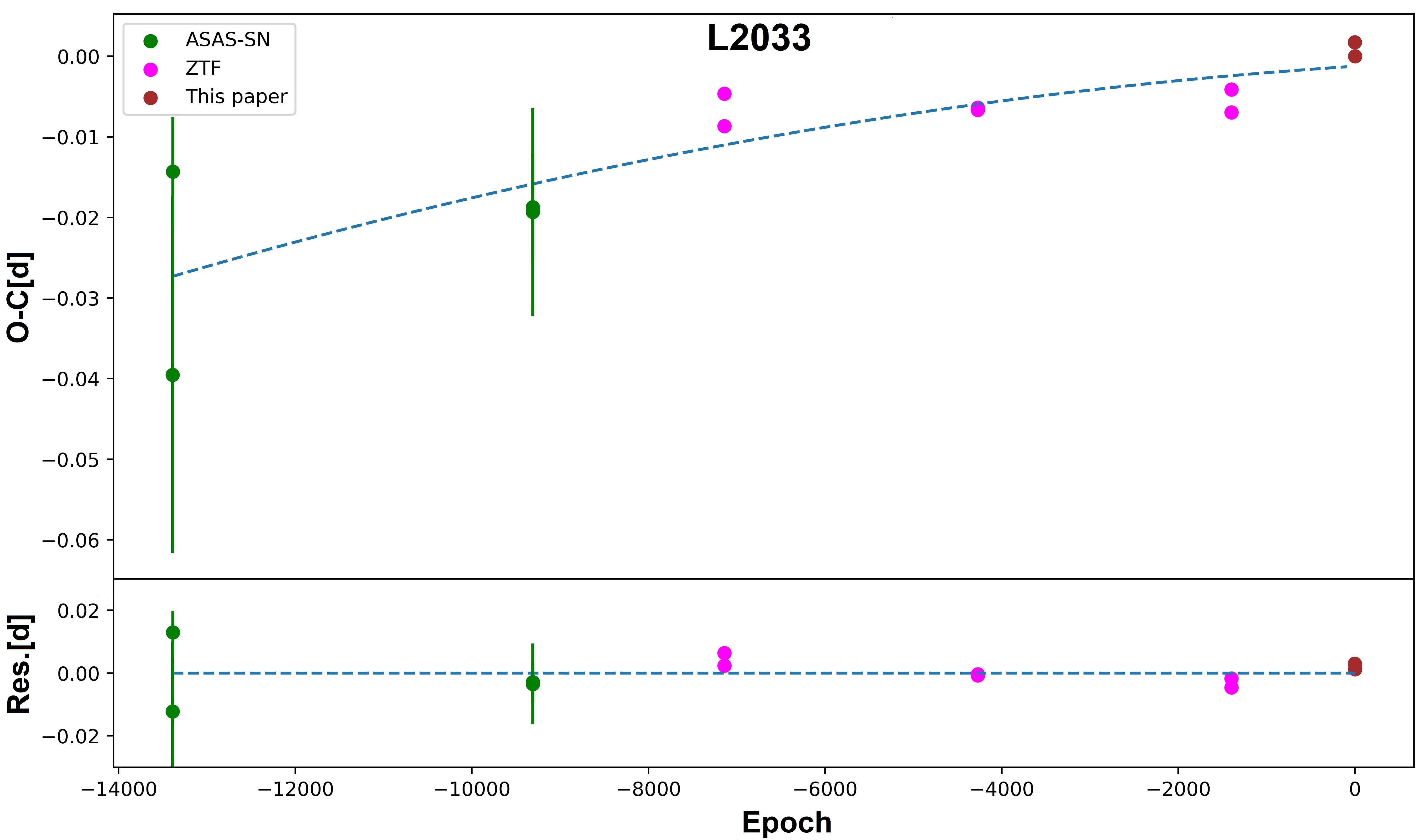}
\includegraphics[width=0.47\textwidth]{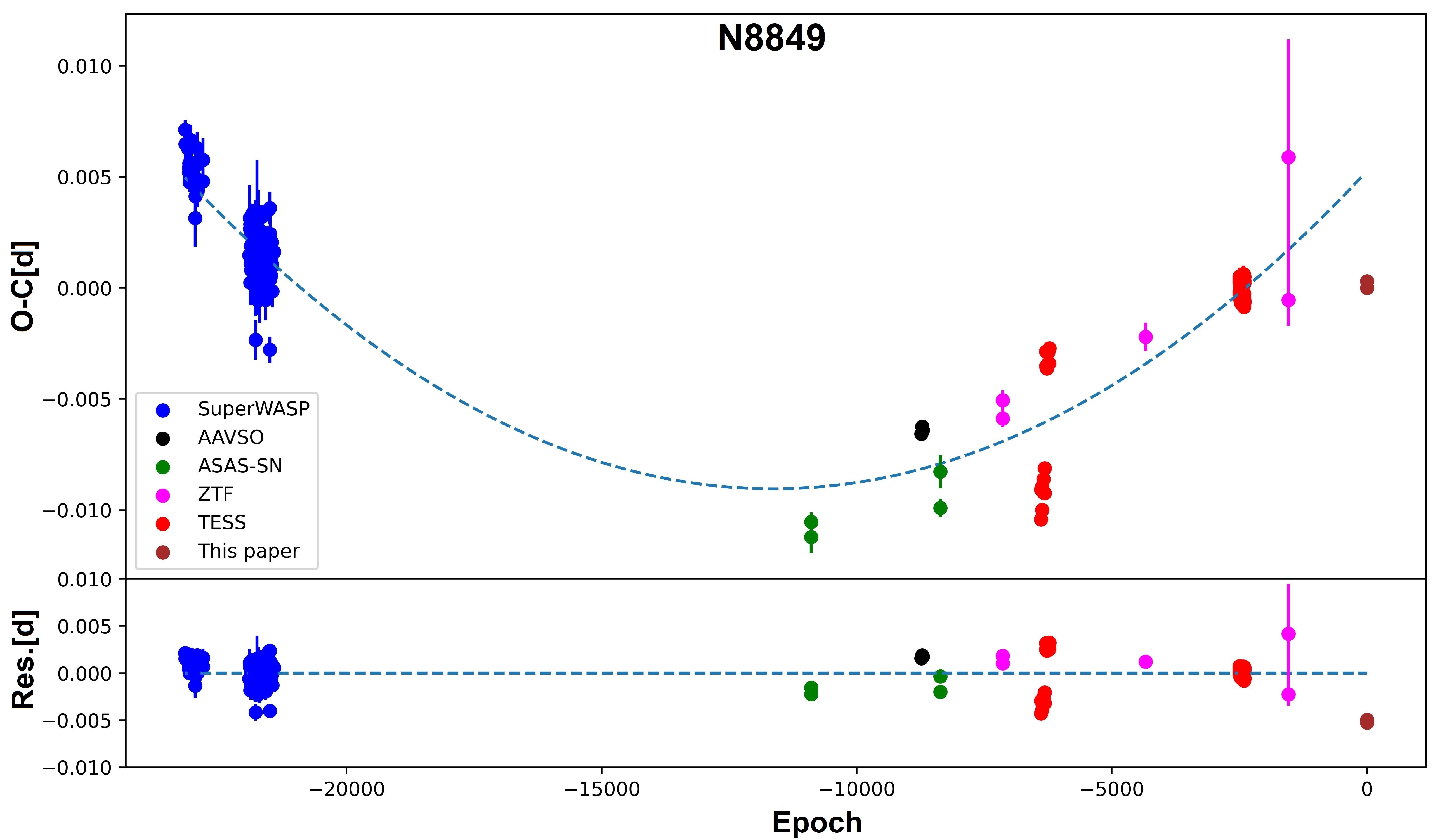}
\includegraphics[width=0.47\textwidth]{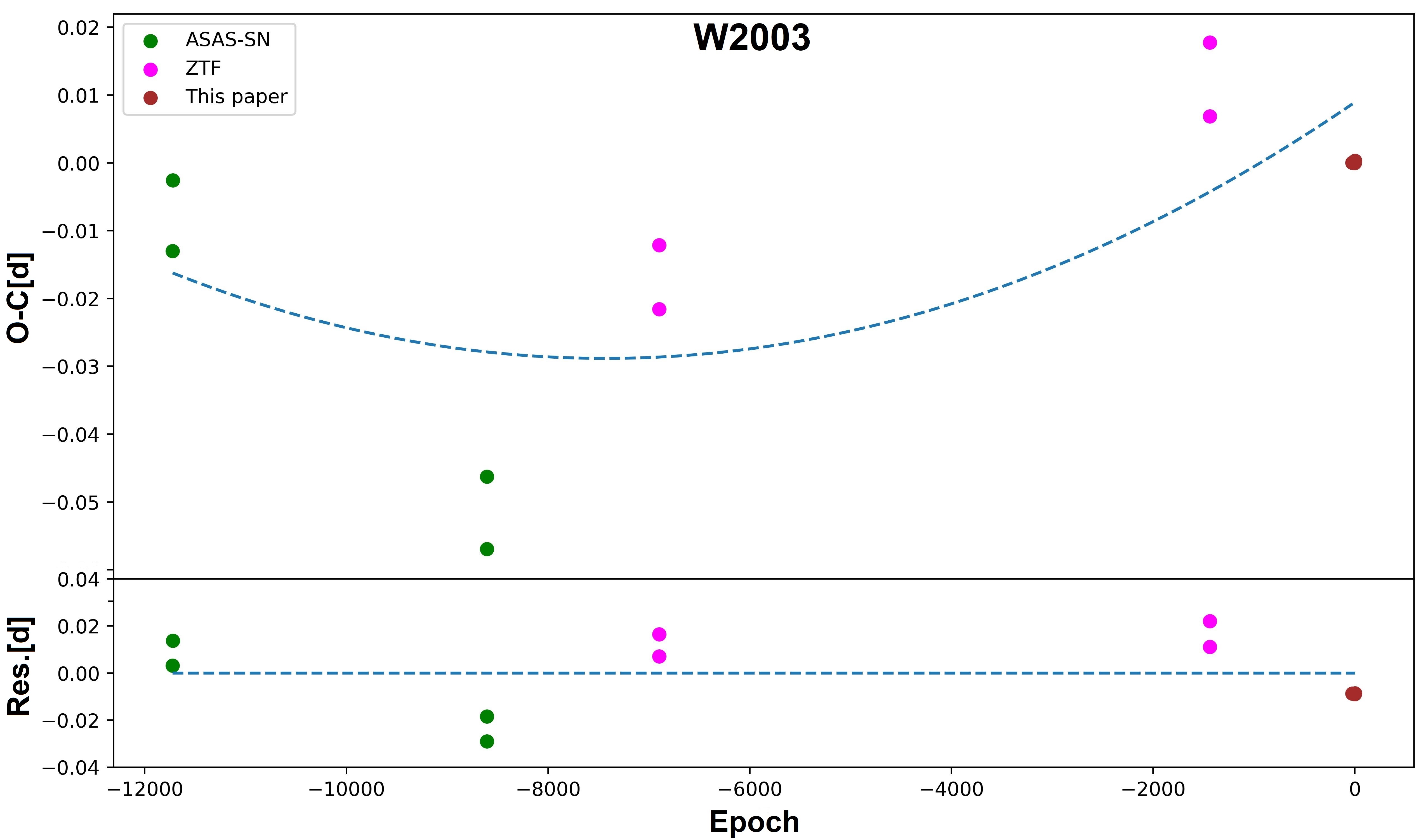}
\includegraphics[width=0.47\textwidth]{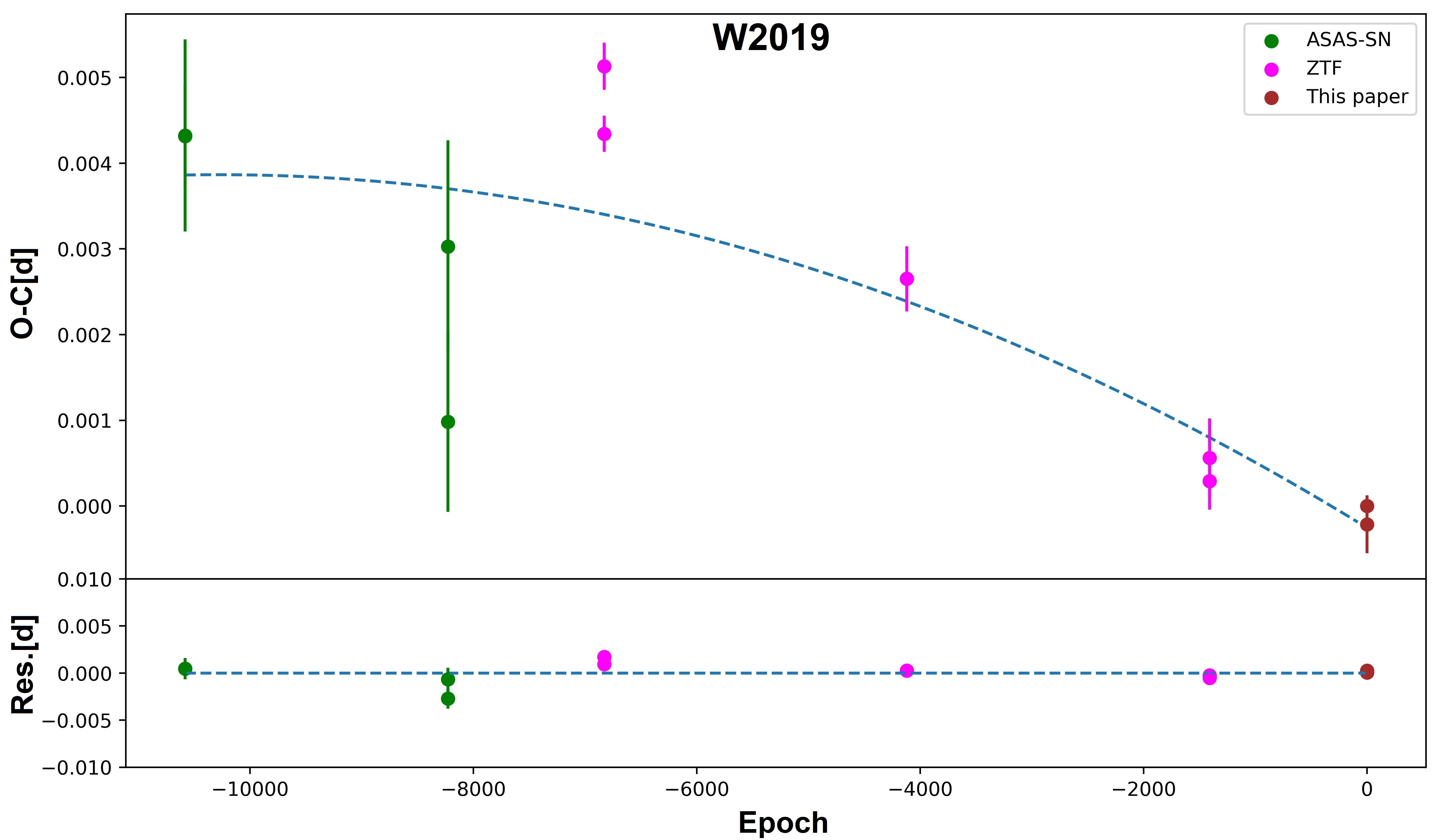}
\caption{The O-C diagrams of the eight targets.}
\label{O-C}
\end{figure*}

To study the orbital period changes of the eight targets, the O-C (observed eclipse timing minus calculated eclipse timing) method \citep{Li2013} was applied. We collected as many eclipse timings as possible by using the data of photometric surveys including All-Sky Automated Survey for SuperNovae \cite[ASAS-SN;][]{shappee2014,jayasinghe2018}, the Zwicky Transient Facility \cite[ZTF;][]{ztf1,ztf2}, the Transiting Exoplanet Survey Satellite \cite[TESS;][]{tess}, Wide Angle Search for Planets \cite[SuperWASP;][]{wasp}, and American Association of Variable Star Observers (AAVSO). For the data of AAVSO, SuperWASP and TESS, we can directly compute the eclipse timings using the \cite{kw1956} method. This method essentially involves fitting a parabola to the data points near the minimum, and the time corresponding to the extremum represents the eclipse timing. While for the data of ASAS-SN and ZTF, we used the period shift method proposed by \cite{Li_2020} to shift the discrete data into one period before calculating the moment of eclipse minimum. Then, we converted all the Heliocentric Julian Date ($HJD$) time to Barycentric Julian Date in Barycentric Dynamical Time ($BJD_{TDB}$) using the online tool\footnote{\url {https://astroutils.astronomy.osu.edu/time/hjd2bjd.html}} developed by \cite{Eastman2010}.  The eclipsing times extracted from our observations are listed in Table \ref{min}. The online machine-readable format is available for the extracted and collected minima times of the target binary systems. The O-C values were calculated using the following reference ephemeris,

\begin{equation}
\begin{aligned}
BJD=BJD_0+P\times E,
\end{aligned}
\end{equation}

\noindent where $BJD$ is the observational eclipse timings, $BJD_0$ listed in the second column of Table \ref{ephemeris} is the initial primary eclipse timing, and $P$ listed in the third column of Table \ref{ephemeris} is the orbital period. The calculated O-C values are presented in Table \ref{min}, and an online machine-readable format is available. The corresponding O-C diagram is displayed in Figure \ref{O-C}. As seen in Figure \ref{O-C}, all targets show long term variations, thus we used the following equation for the O-C fitting,

\begin{equation}
\begin{aligned}
O-C=\Delta{T_0}+\Delta{P_0}\times E+\frac{\beta}{2}{E^2}.
\end{aligned}
\end{equation}

We performed the fitting using the least squares method, with the parameter errors calculated from their covariance matrix. The determined parameters are shown in Table \ref{tabO-C}, and the corrected new ephemerides are listed in Table \ref{ephemeris}.
Four stars exhibit long-term increase orbital period, while the others show long-term decrease orbital period.

\renewcommand\arraystretch{1.2}
\begin{table*}
\caption{The times of minima extracted from our ground-based observations.}
\centering
\small
\begin{tabular}{c c c c c}
\hline
System & Min.($BJD_{TDB}$) & Error & Epoch & O-C\\ 
\hline
J1655	&2460462.7943	&0.0003	&	0	&	0		\\
	&2460462.9281	&0.0003	&	0.5	&-0.0012		\\
J1708	&2460492.7521	&0.0002	&	-0.5	&0.0011	\\
	&2460492.8925	&	0.0002	&	0	&	0		\\
J1709	&2460491.7594	&	0.0003	&	0	&	0		\\
	&	2460491.8980	&	0.0002	&	0.5	&	-0.0003		\\
J2249	&	2460559.7401	&	0.0002	&	0	&	0		\\
	&	2460559.8971	&	0.0002	&	0.5	&	-0.0005		\\
L2033	&	2460489.7876	&	0.0003	&	-0.5	&	0.0017		\\
	&	2460489.9251	&	0.0002	&	0	&	0		\\
N8849	&	2460539.6833	&	0.0002	&	-0.5	&	0.0003		\\
	&	2460539.8259	&	0.0002	&	0	&	0		\\
W2003	&	2460513.9133	&	0.0002	&	-26.5	&	0		\\
	&	2460521.6774	&	0.0002	&	0	&	0		\\
	&2460521.8242	&0.0002	&	0.5	&	0.0003	\\
W2019	&	2460515.8006	&	0.0002 &	0	&	0		\\
	&2460515.9479	&	0.0003	&	0.5	&	-0.0002	\\
\hline
\end{tabular}
\label{min}
\end{table*}

\renewcommand\arraystretch{1.2}
\begin{table*}
\caption{\centering Reference and new ephemeris of the systems. The reference times of minimum ($t_0$) were obtained from our observations in this study.}
\centering
\small
\begin{tabular}{c|cc|cc}
\hline
System& \multicolumn{2}{c|}{Reference ephemeris}& \multicolumn{2}{c}{New ephemeris}\\ 
&$t_0(BJD_{TDB})$&Period(day)/Source& Corrected $t_0(BJD_{TDB})$&New Period(day)\\ 
\hline
J1655 &  2460462.7942(3)&0.2699918/ASAS-SN& 2460462.793(3)&0.2699974(9)\\ 
J1708 &  2460492.8925(2)&0.2829928/ASAS-SN& 2460492.893(1)&0.2829954(6)\\ 
J1709 &  2460491.7594(3)&0.2777515/ASAS-SN& 2460491.759(1)&0.2777499(4)\\ 
J2249 &  2460559.7401(2)&0.3150597/ASAS-SN& 2460559.740(2)&0.3150594(5)\\ 
L2033 &  2460489.9251(2)&0.2785969/ASAS-SN& 2460489.924(1)&0.278598(3)\\
 N8849 & 2460539.8259(2)& 0.2857171/ASAS-SN& 2460539.832(2)&0.2857196(2)\\
 W2003 & 2460521.6774(2)& 0.2929848/ASAS-SN& 2460521.686(2)&0.292995(9)\\
 W2019 & 2460515.8006(2)& 0.2951124/ASAS-SN& 2460515.800(2)& 0.2951116(8)\\
 \hline
\end{tabular}
\label{ephemeris}
\end{table*}

\renewcommand\arraystretch{1.2}
\begin{table*}
\caption{\centering The O-C fitting coefficients and mass transfer rate.}
\centering
\small
\begin{tabular}{ccccccccc}
\hline
Parameter& $\Delta{T_0}$& Error& $\Delta{P_0}$& Error& $\beta$&Error & $dM_1/dt$&Error\\ 
&$(\times {10^{-4}} d)$&& $(\times {10^{-7}} d)$&& $(\times {10^{-7}} d$ $ {yr^{-1}})$& & $(\times {10^{-7}} M_\odot$ $ {yr^{-1}})$&\\ 
\hline
J1655 &  -10 & 27 & 56 & 11 & 5 & 2 & -3 & 1\\
J1708 &  3 & 13 & 26 & 6 & 2 & 1 & 4 & 2\\
J1709 & 0 & 10 & -16 & 4 & -2 & 1 & 4 & 2\\
J2249 & -2 & 22 & -3 & 5 & -1 & 1 & 2 & 1\\
L2033 & -12 & 78 & 7 & 29 & -2 & 6 & 2 & 6\\
N8849 & 53 & 4 & 25 & 2 & 6 & 1 & 15 & 1\\
\hline
\end{tabular}
\label{tabO-C}
\end{table*}

\vspace{0.6cm}
\section{Light Curve Solutions}
\label{sec5}
Version 2.4.9 of the PHysics Of Eclipsing BinariEs (PHOEBE) Python code was utilized to analyze the light curves of the target binary systems (\citealt{2016ApJS..227...29P}, \citealt{2020ApJS..250...34C}). We converted time to phase in the light curve using a new ephemeris (Table \ref{ephemeris}). The contact mode was chosen for light curve modeling due to the observed light curve shapes, the systems' catalog classifications, and their short orbital periods. The gravity-darkening coefficient was set to $g_1=g_2=0.32$ (\citealt{1967ZA.....65...89L}), and the bolometric albedo to $A_1=A_2=0.5$ (\citealt{1969AcA....19..245R}). The stellar atmosphere model was adopted from the study by \cite{2004AA...419..725C}, and the limb darkening coefficients were treated as free parameters in PHOEBE.

The initial effective temperature ($T$) was first taken from the Gaia DR3 database for the analysis, with the TIC value used if Gaia DR3 did not report it (Table \ref{systemsinfo}). This temperature was assigned based on the depth of the minima in the light curves, corresponding to the hotter component of the systems. The initial effective temperature of the cooler star was estimated based on the difference in depth between the primary and secondary minima in the light curves.

\begin{figure*}
\centering
\includegraphics[width=0.31\textwidth]{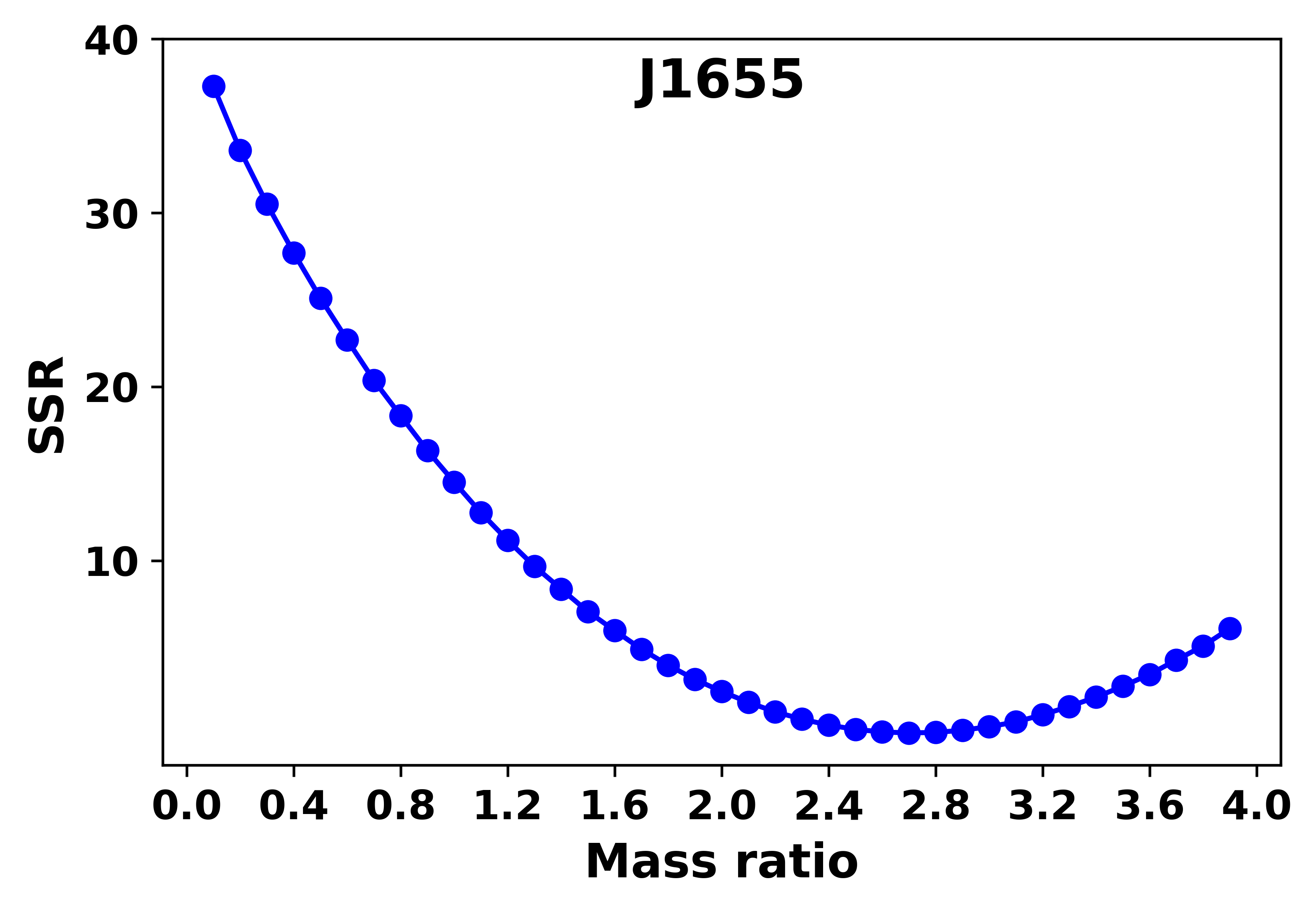}
\includegraphics[width=0.31\textwidth]{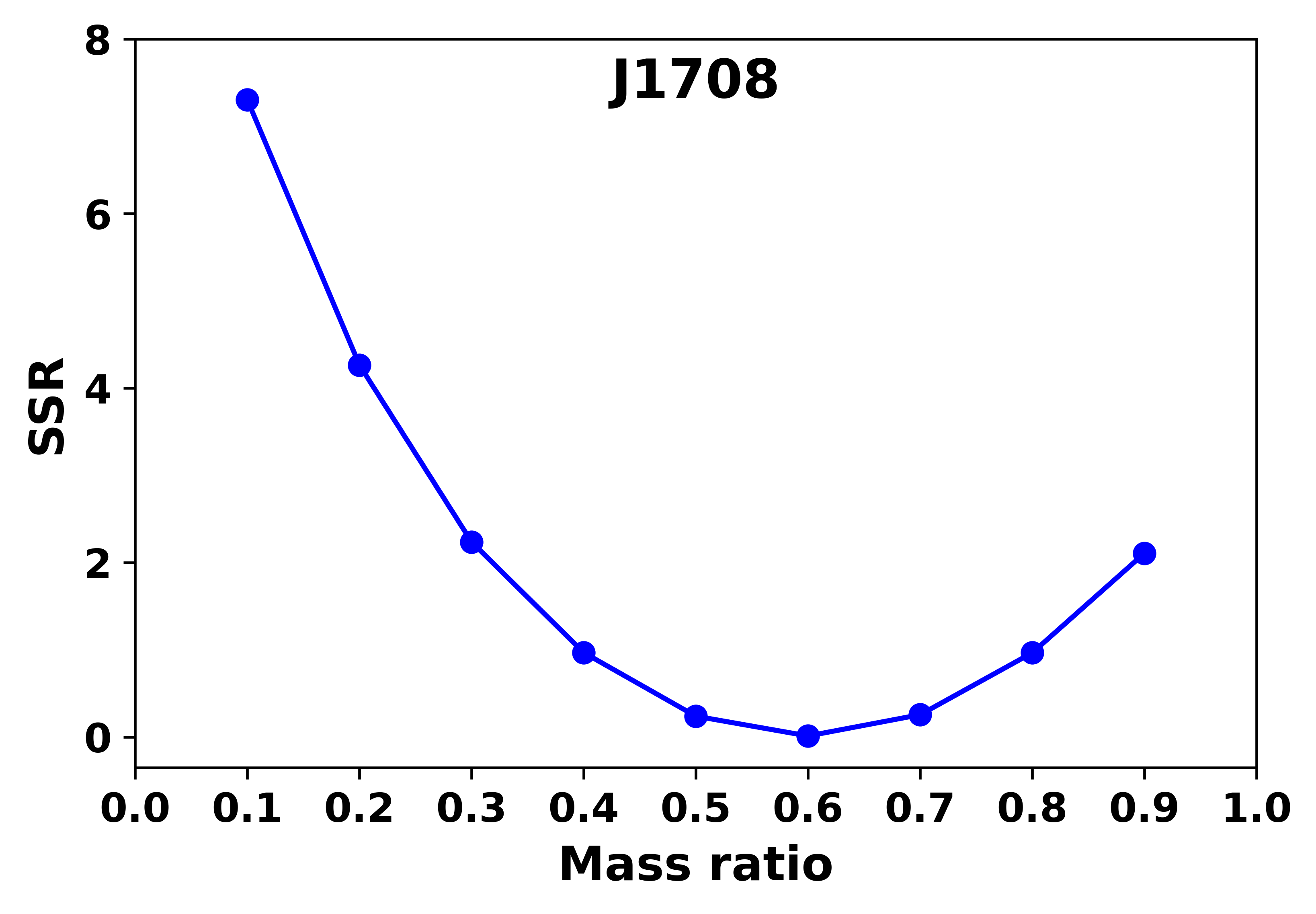}
\includegraphics[width=0.31\textwidth]{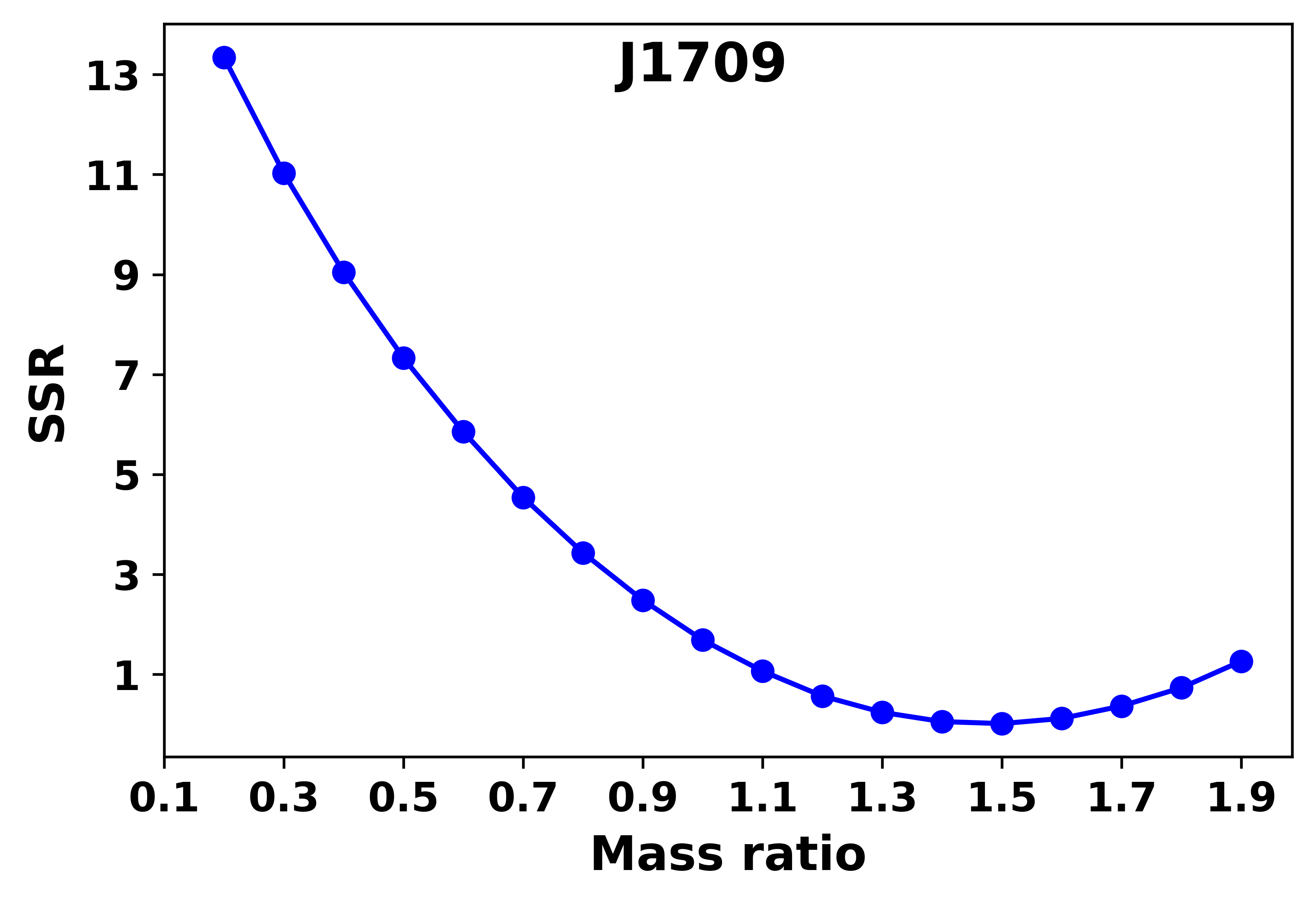}
\includegraphics[width=0.31\textwidth]{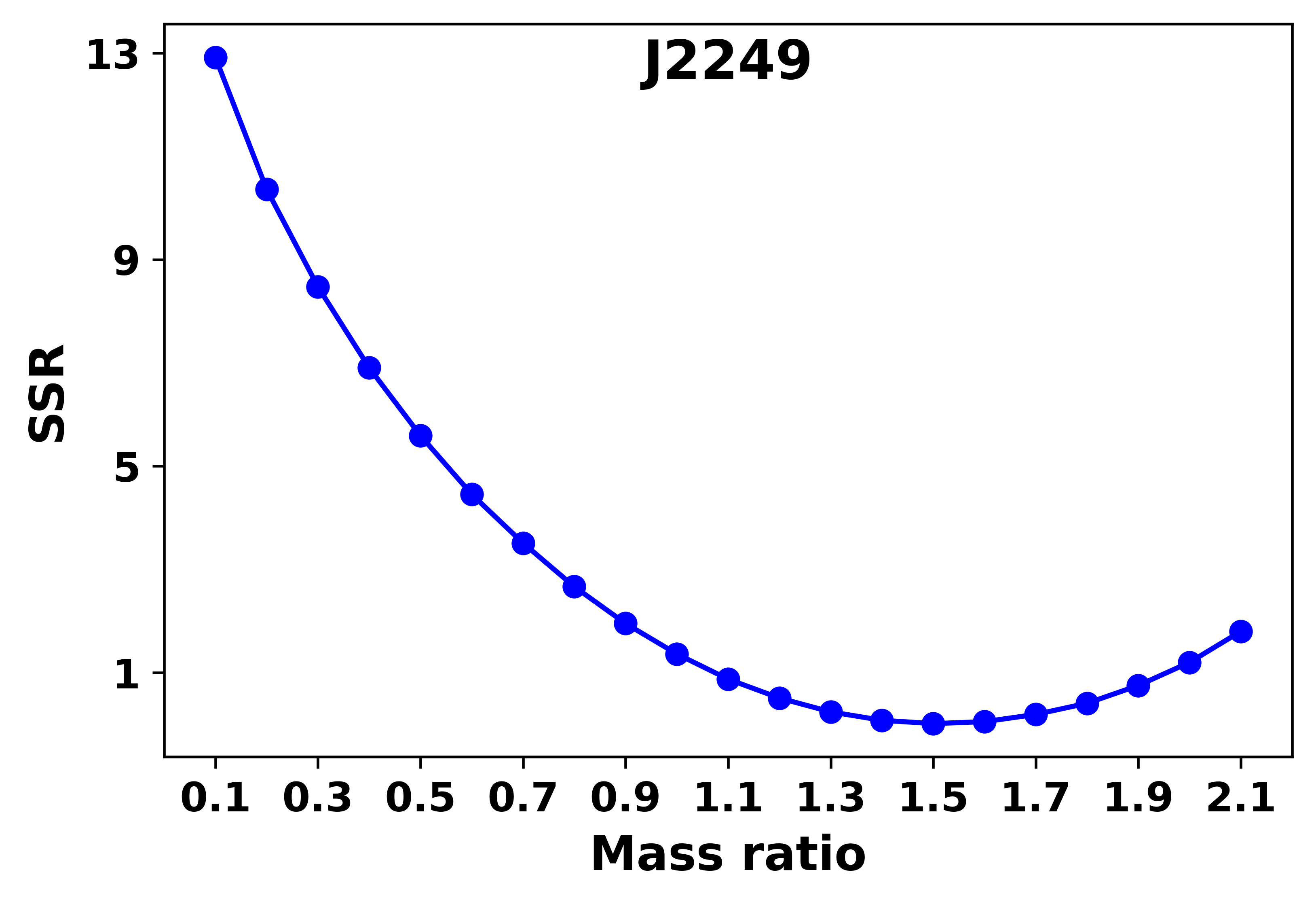}
\includegraphics[width=0.31\textwidth]{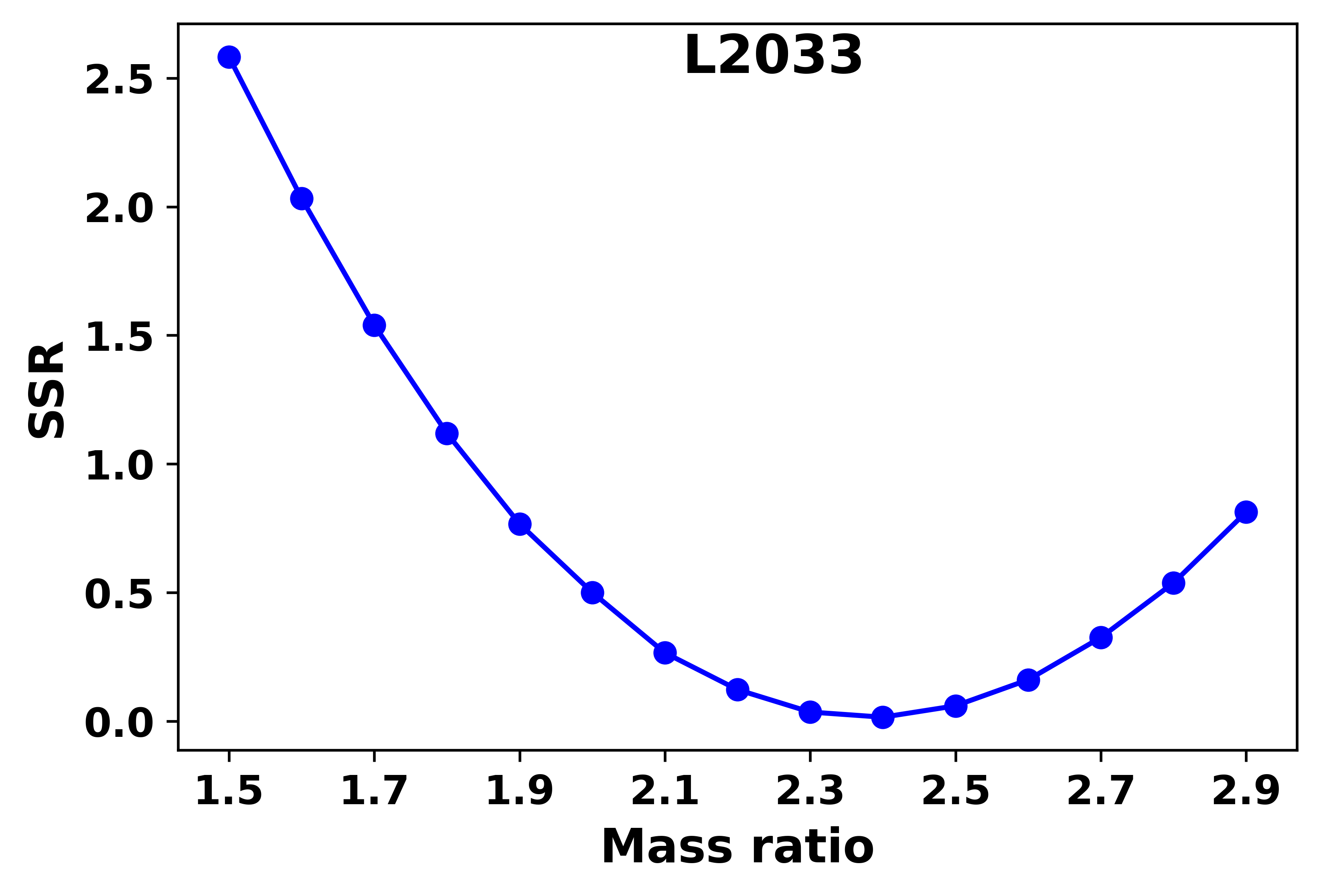}
\includegraphics[width=0.31\textwidth]{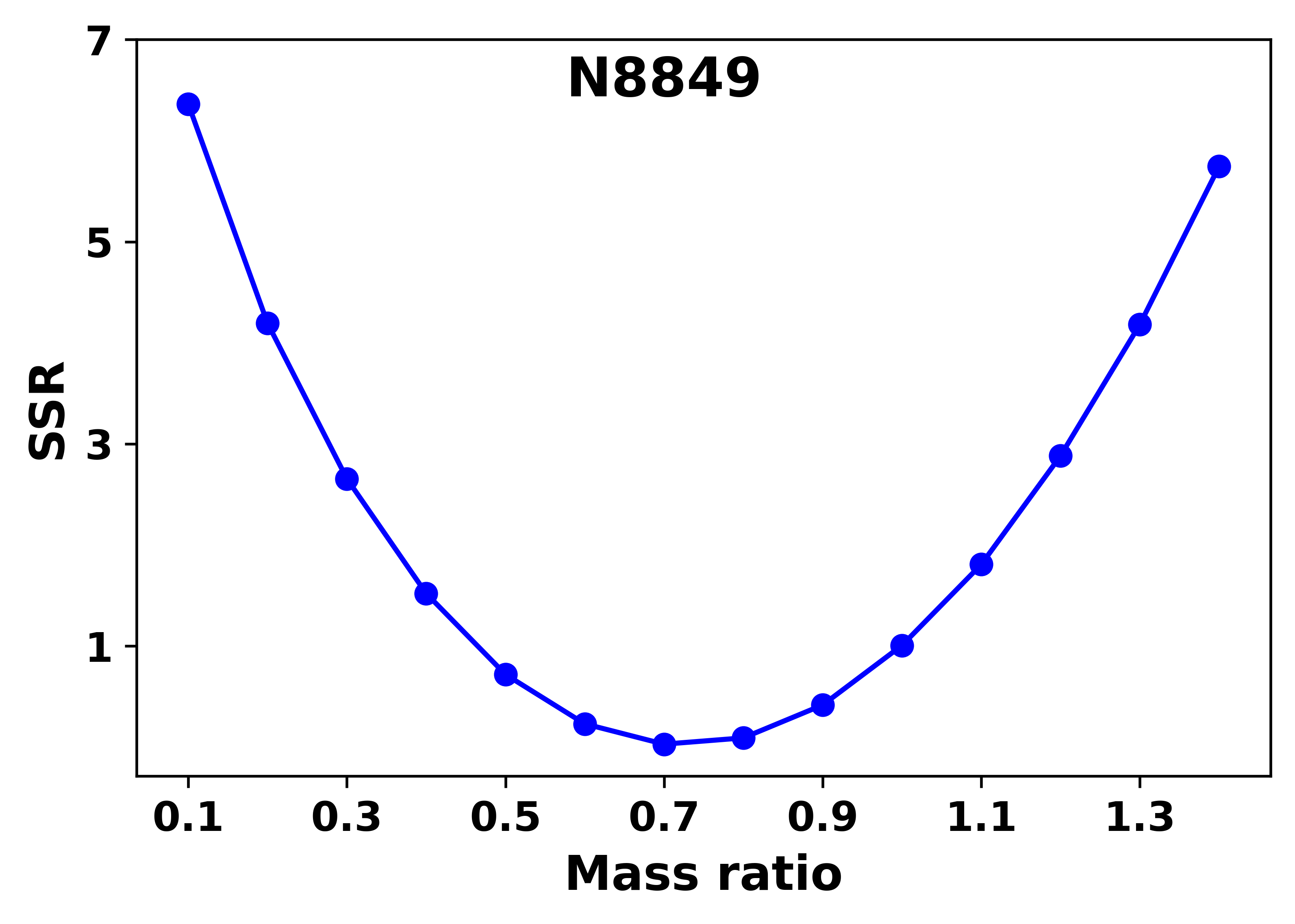}
\includegraphics[width=0.31\textwidth]{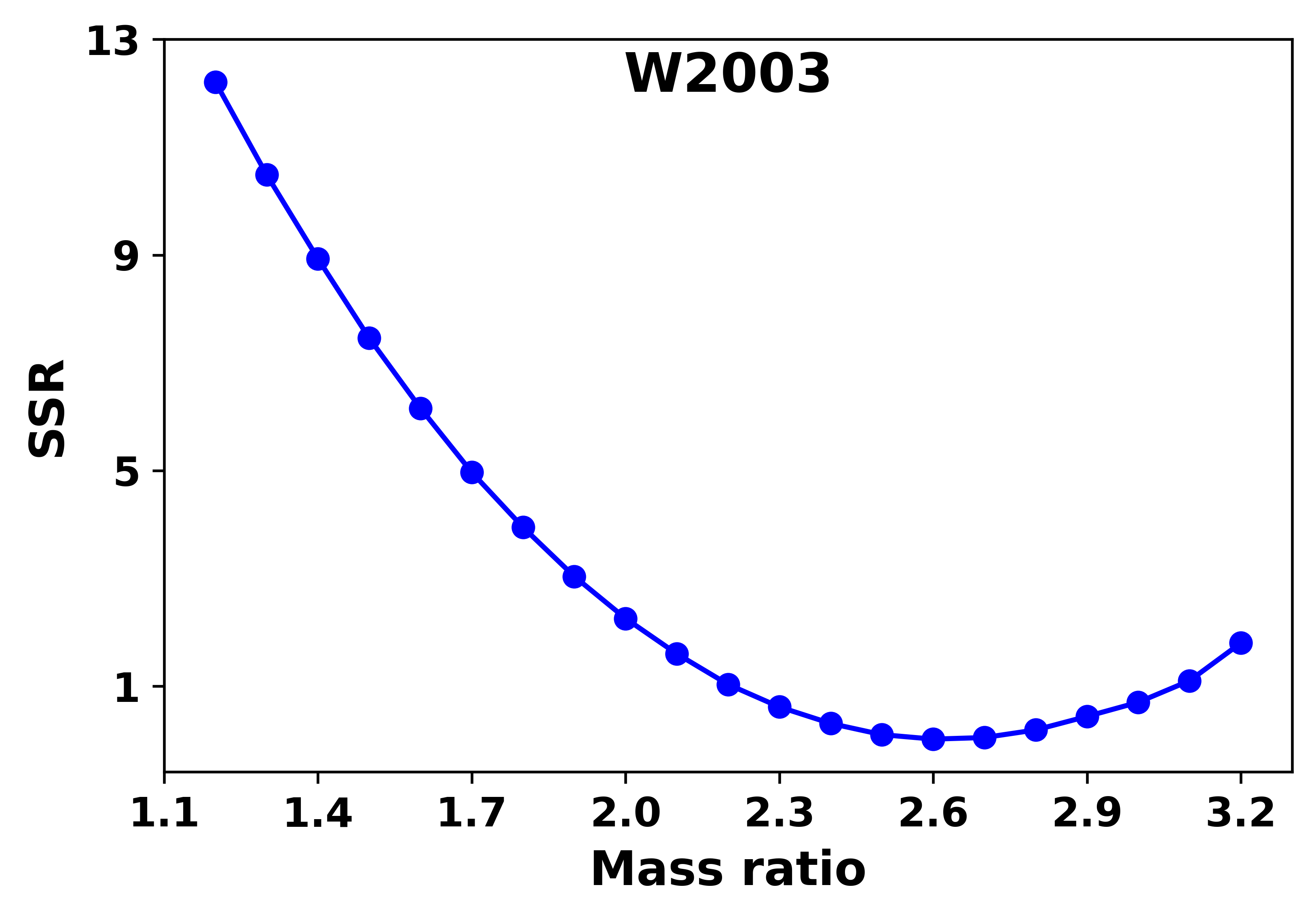}
\includegraphics[width=0.31\textwidth]{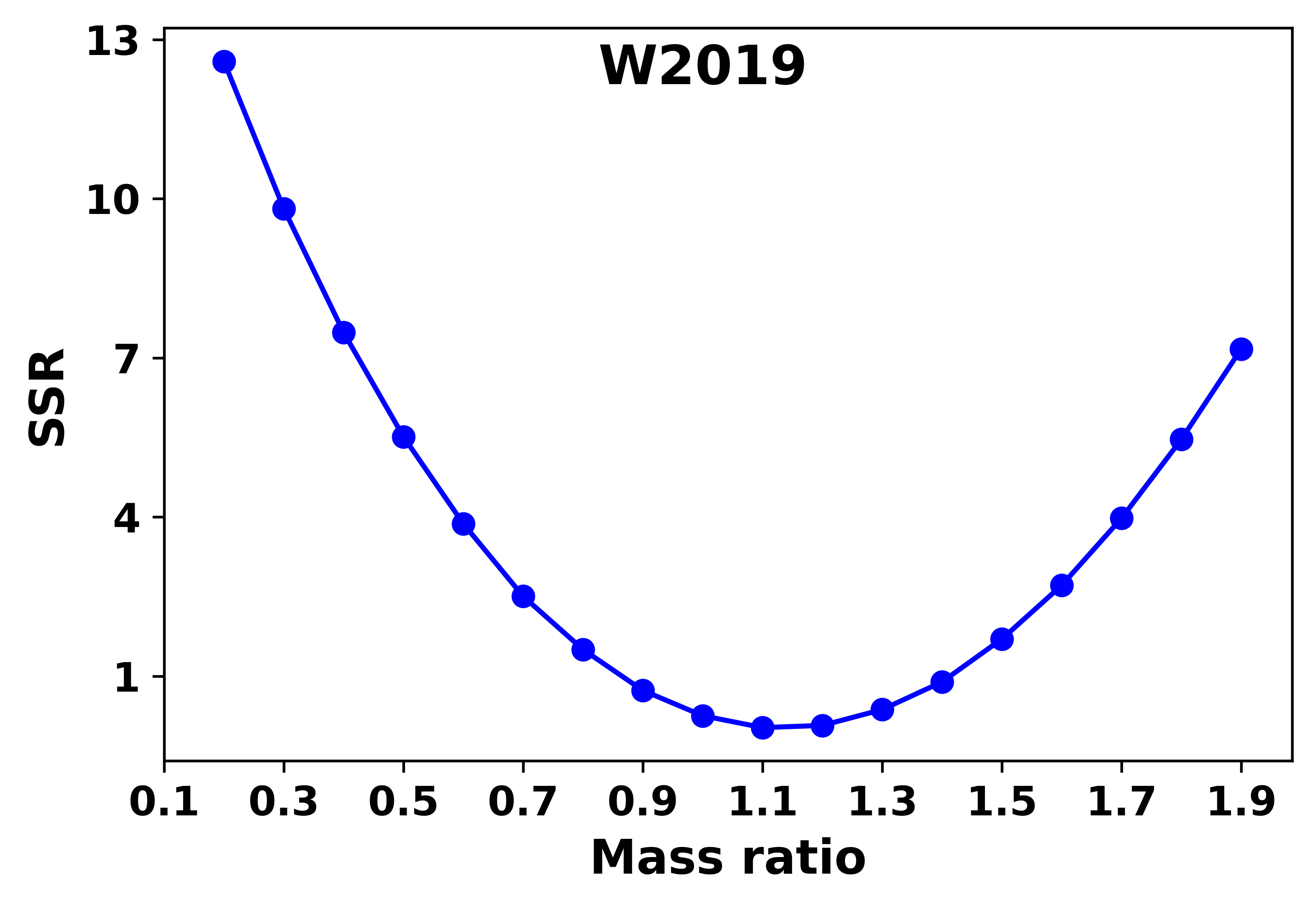}
\caption{The sum of squared residuals as a function of mass ratio.}
\label{q-diagrams}
\end{figure*}

The initial mass ratio of the systems was determined using the $q$-search method (\citealt{2005ApSS.296..221T}). Mass ratio values ranging from 0.1 to 12 were explored for all target systems. Then, a narrower interval was explored to refine the estimate, minimizing the sum of squared residuals. Figure \ref{q-diagrams} demonstrates that each $q$-search curve has a clear minimum sum of squared residuals. According to numerous studies, such as \cite{2021AJ....162...13L} and \cite{2024AJ....168..272P}, the mass ratio in systems with high orbital inclination and total eclipses provides reliable values.

The analysis indicates that J1708 and L2033 systems exhibit unequal maxima in their light curves. Therefore, a cold starspot was needed on one of the components to explain the asymmetry in the light curve maxima. This phenomenon is most plausibly explained by the magnetic activity of the stars, giving rise to the presence of starspots and is known as the O'Connell effect (\citealt{1951PRCO....2...85O}, \citealt{2017AJ....153..231S}). The BSN application version 1.0 (\citealt{paki2025bsn}) was employed to determine the location of the stellar spot. Leveraging its interactive 3D visualization and rotation features, the software offers precise tools for identifying the spot's position. Once established, the spot parameters were incorporated into the PHOEBE code. The inclusion and placement of the stellar spot led to an improved agreement between the synthetic light curve and the observational data.

First, the light curve modeling was performed in PHOEBE using photometric multiband data and initial parameter values to establish the preliminary solutions and achieve a satisfactory theoretical fit. Then, PHOEBE's built-in optimization tool was utilized to further refine the model parameters and improve the fit, employing the Nelder-Mead (Simplex) algorithm (\citealt{nelder1965simplex}), which operates efficiently without requiring derivative information. In this process, five main parameters—$T_{1,2}$, $q$, $f$, and $i$—along with four parameters describing the starspot were simultaneously optimized to achieve the best agreement between the synthetic and observed light curves.

Given that the modeling and optimization procedures in PHOEBE do not provide uncertainty estimates, we employed the BSN application version 1.0 for this purpose. This application is currently available for Windows OS and is accessible to members of the BSN project. The BSN application generates synthetic light curves over 40 times faster than PHOEBE during the Markov chain Monte Carlo (MCMC) fitting process. This enhanced computational speed is attributed to its optimized architecture and integration of updated computational tools, while the core scientific procedures for light curve analysis remain consistent with those implemented in other binary star modeling software. The results obtained from BSN and PHOEBE, as well as the synthetic light curves generated by both, are highly consistent. The minor discrepancies between them can be ascribed to the differing approaches adopted by each code in modeling the geometrical structure of binary systems, which may naturally lead to slight variations in the resulting synthetic light curves. Accordingly, the MCMC analysis was performed using the BSN application with 24 walkers and 1000 iterations, focusing on five primary parameters ($T_{1,2}$, $q$, $f$, and $i$), resulting in estimates of the parameter uncertainties. The mean upper and lower bounds of these uncertainties are provided in Table \ref{lc-analysis}. It is worth noting that the final results and the synthetic light curves generated by the BSN application and PHOEBE for the target systems were essentially the same.

The analysis revealed that none of the target systems displayed a detectable third light component ($l_3$). The light curves of these systems were successfully modeled without including $l_3$, and there is no indication of contaminating light from nearby stars. Furthermore, our analysis of the orbital period variations did not reveal any evidence suggesting the presence of a third body. Although these results provide strong constraints against the existence of a third light contribution, they cannot definitively rule it out.

\begin{figure*}
\centering
\includegraphics[width=0.495\textwidth]{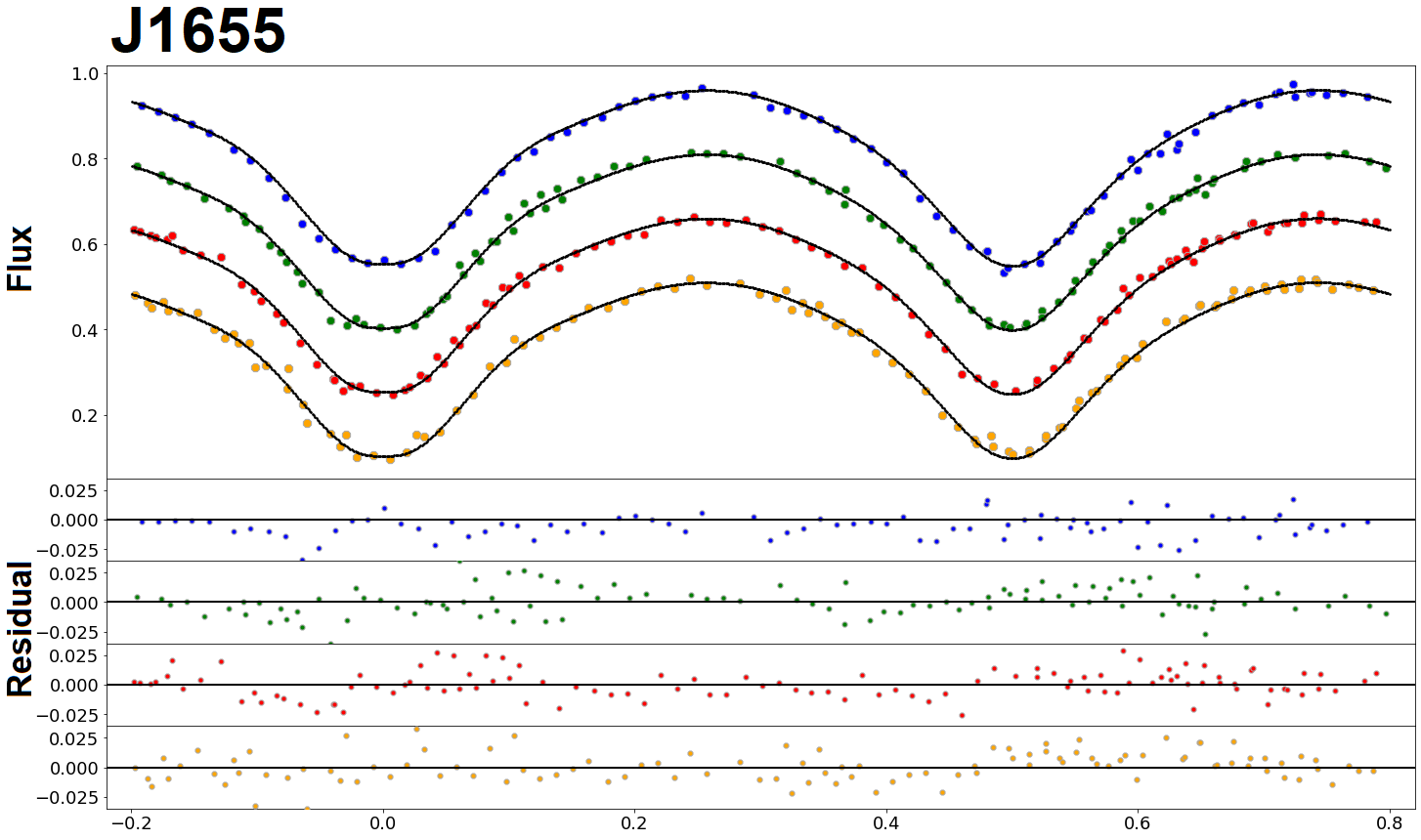}
\includegraphics[width=0.495\textwidth]{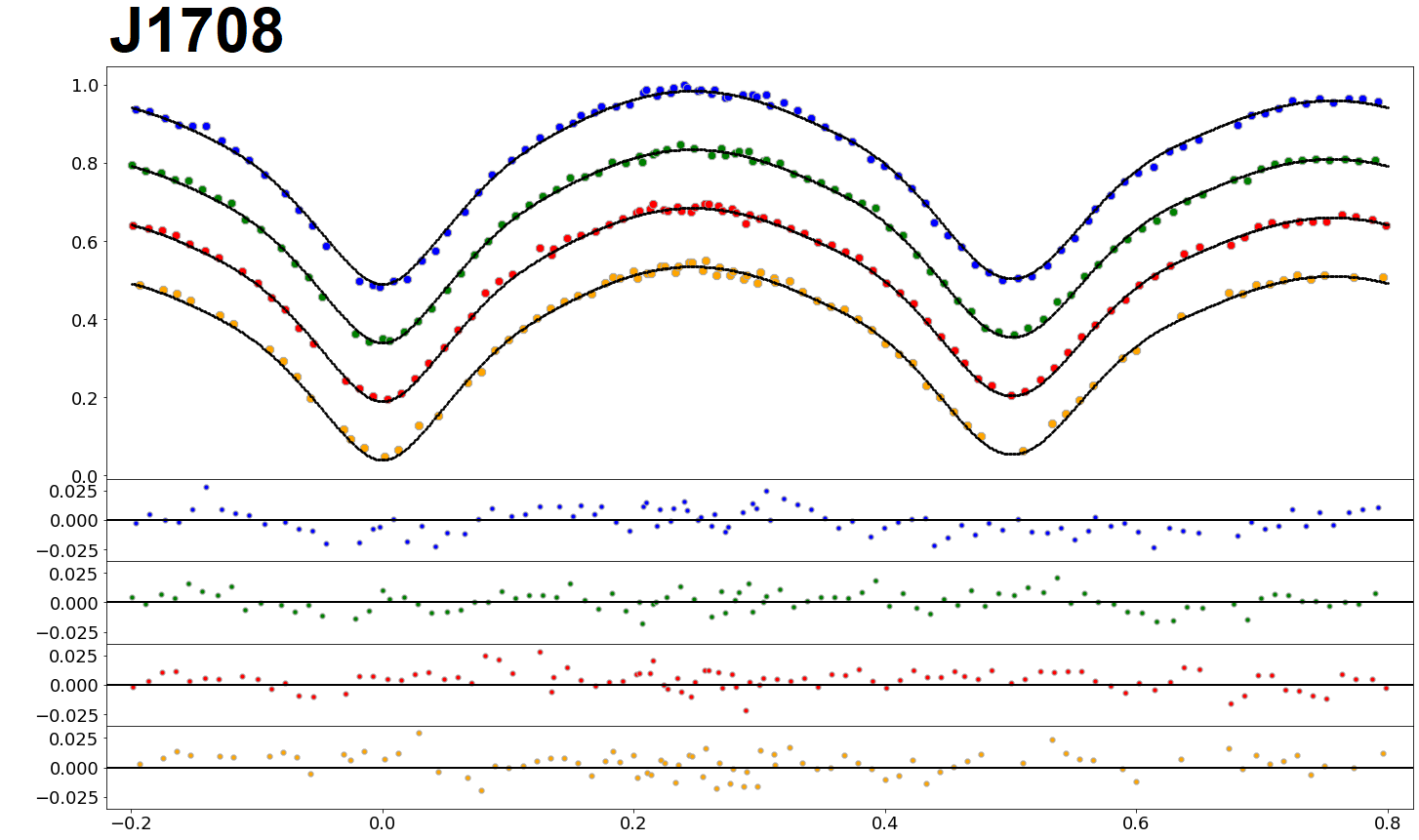}
\includegraphics[width=0.495\textwidth]{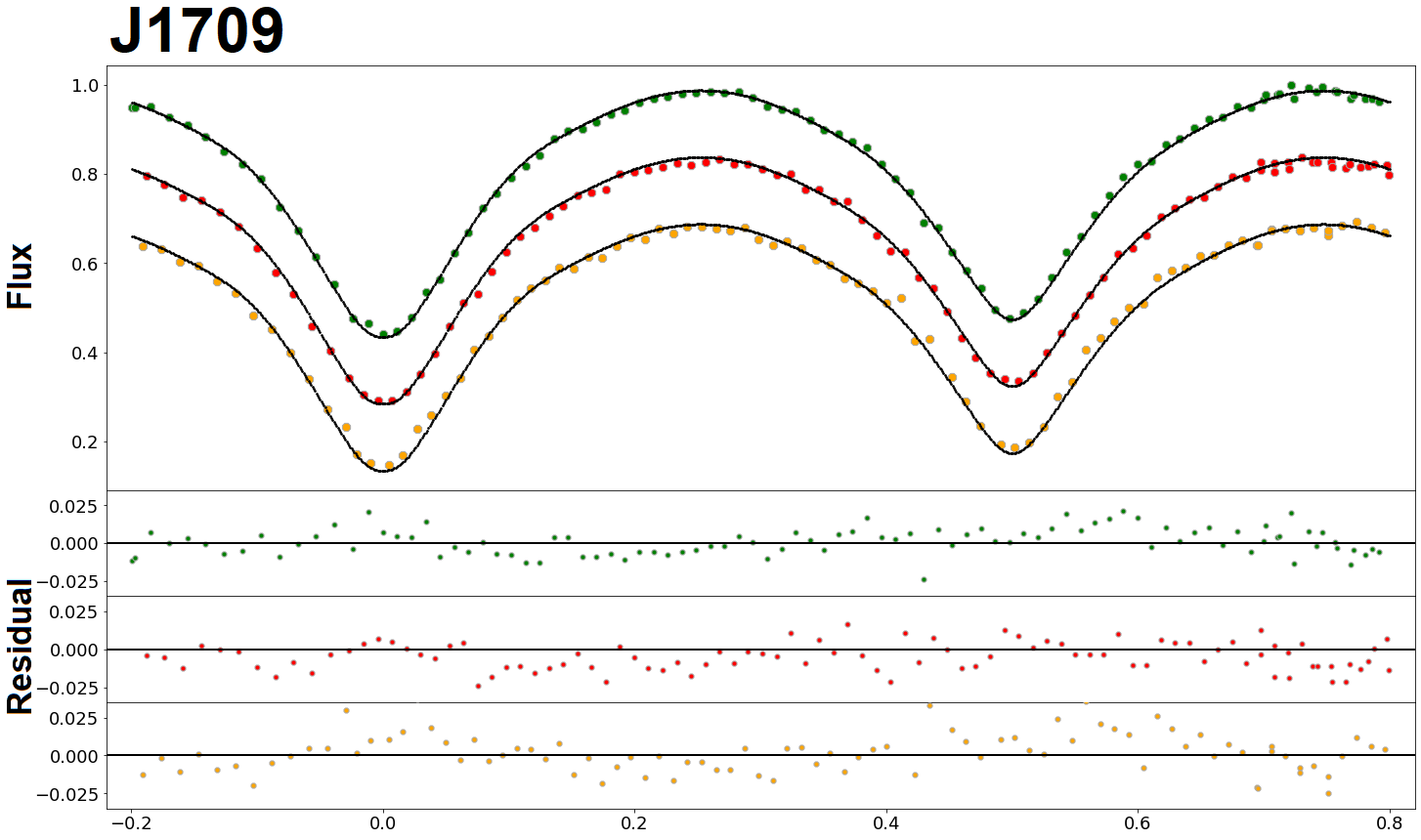}
\includegraphics[width=0.495\textwidth]{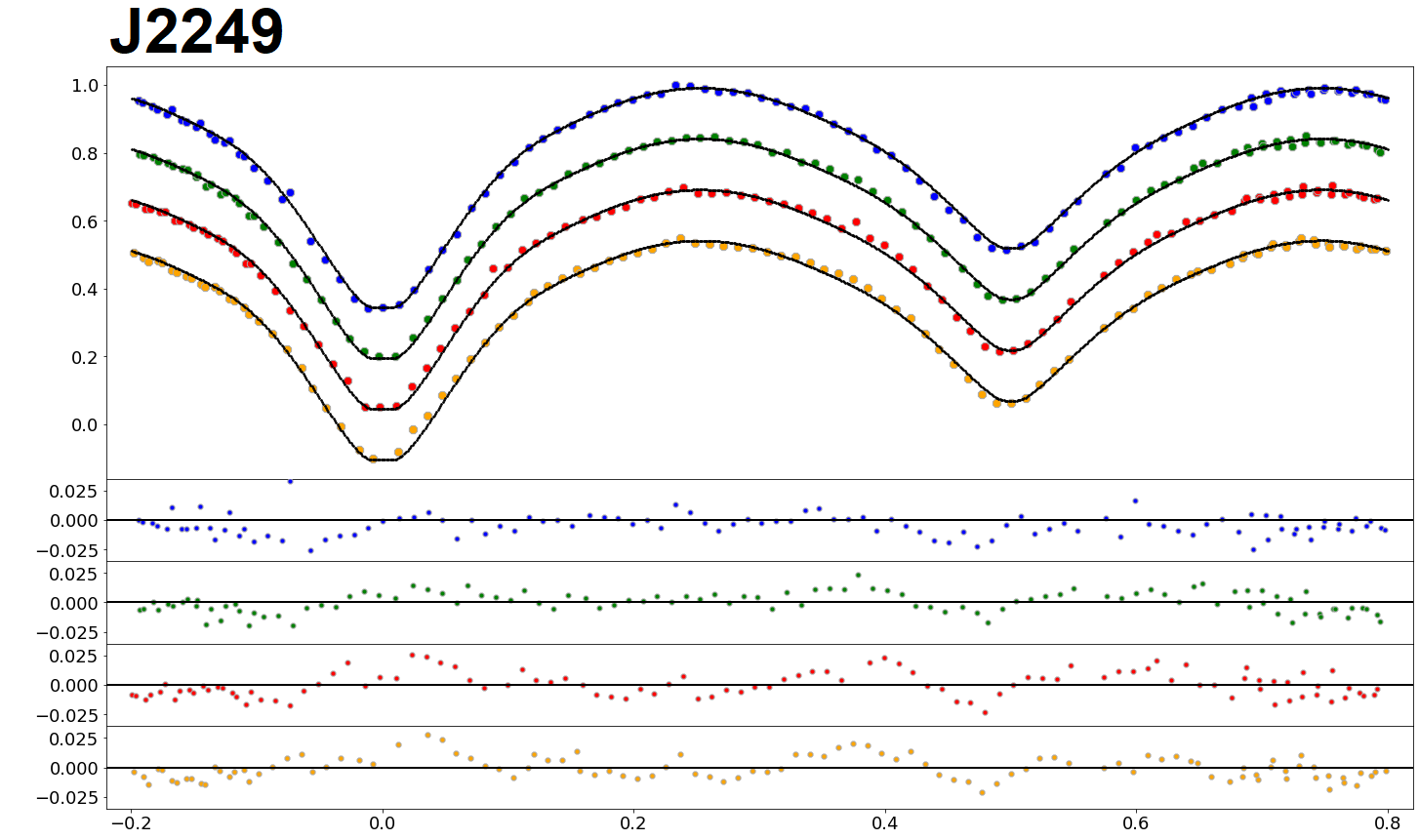}
\includegraphics[width=0.495\textwidth]{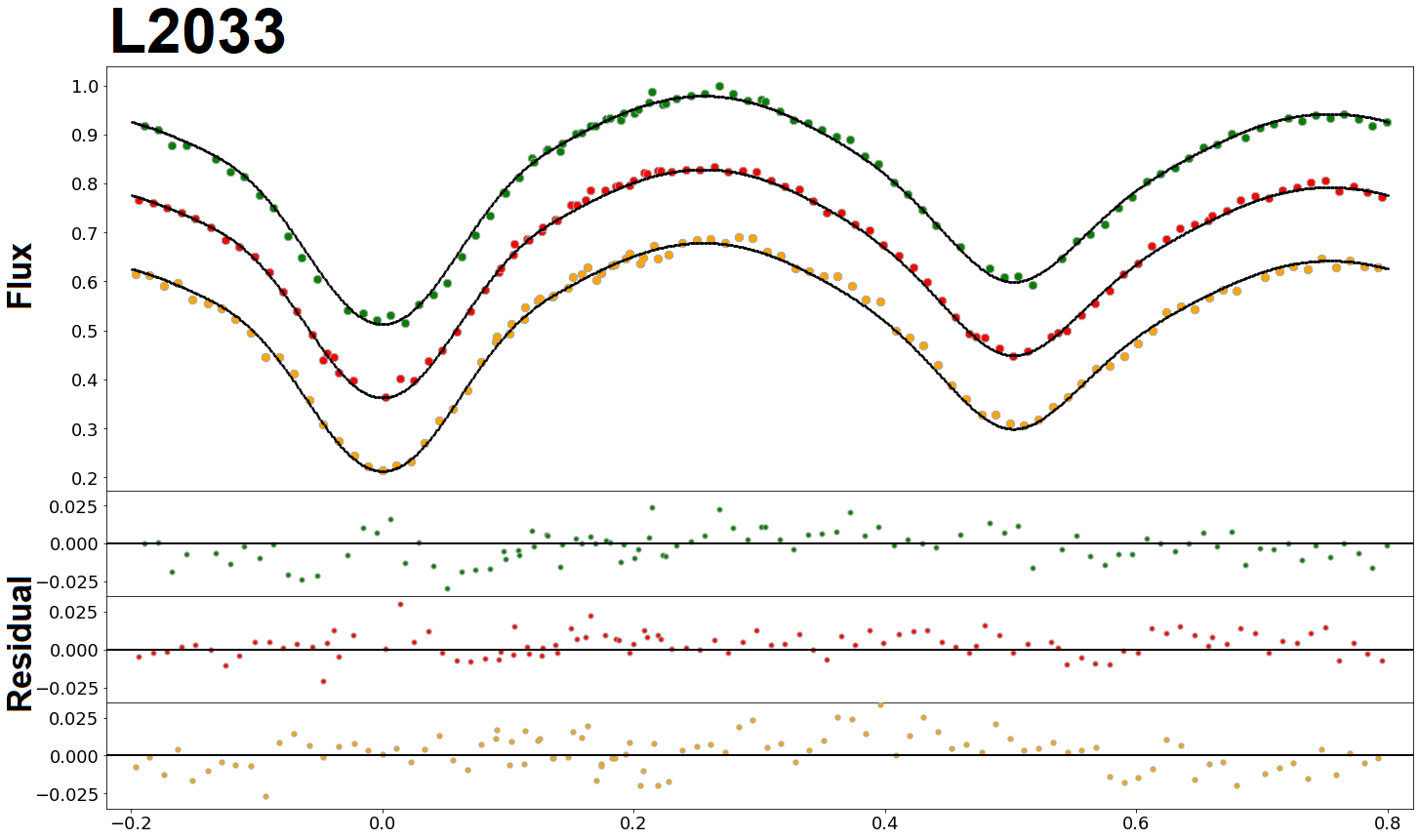}
\includegraphics[width=0.495\textwidth]{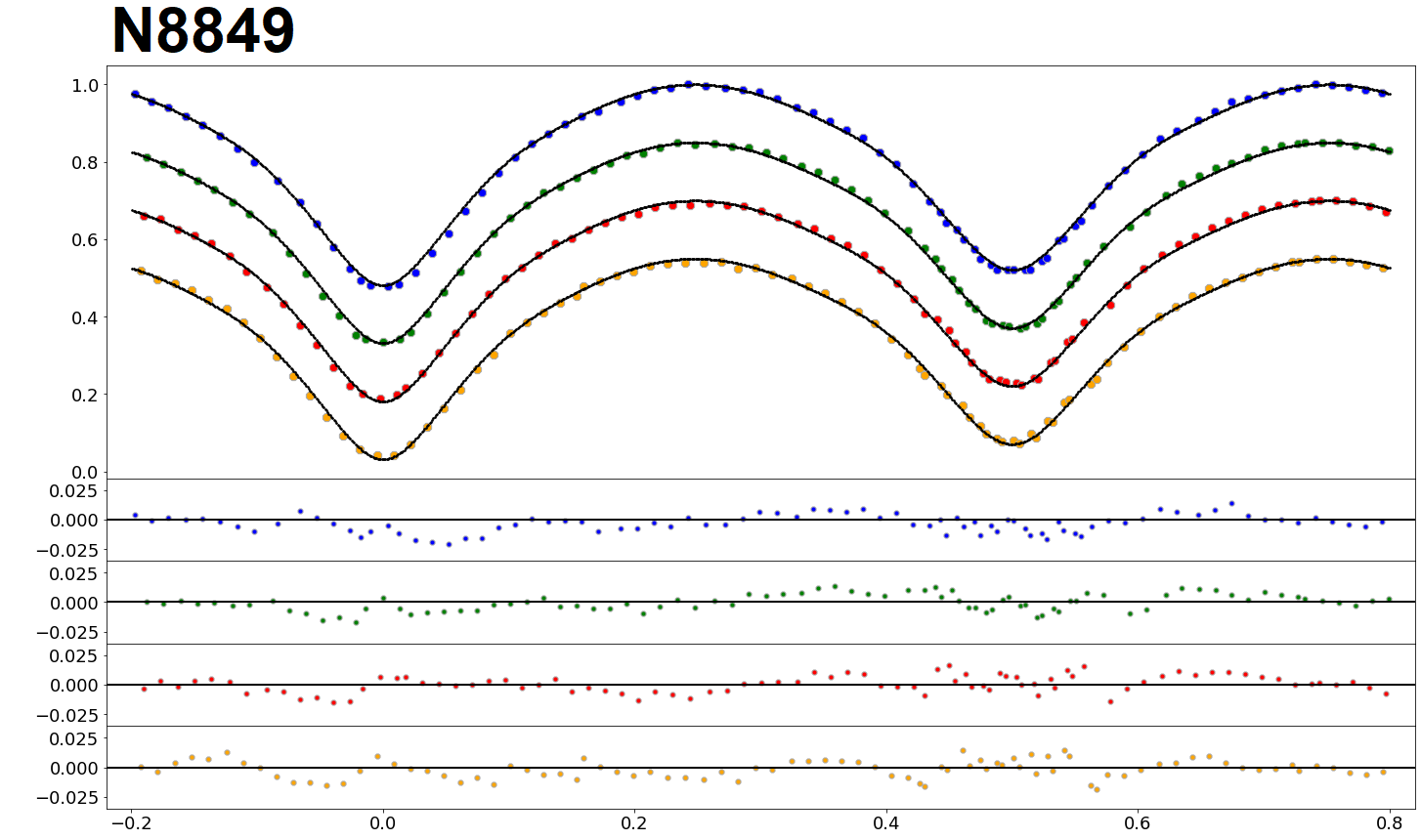}
\includegraphics[width=0.495\textwidth]{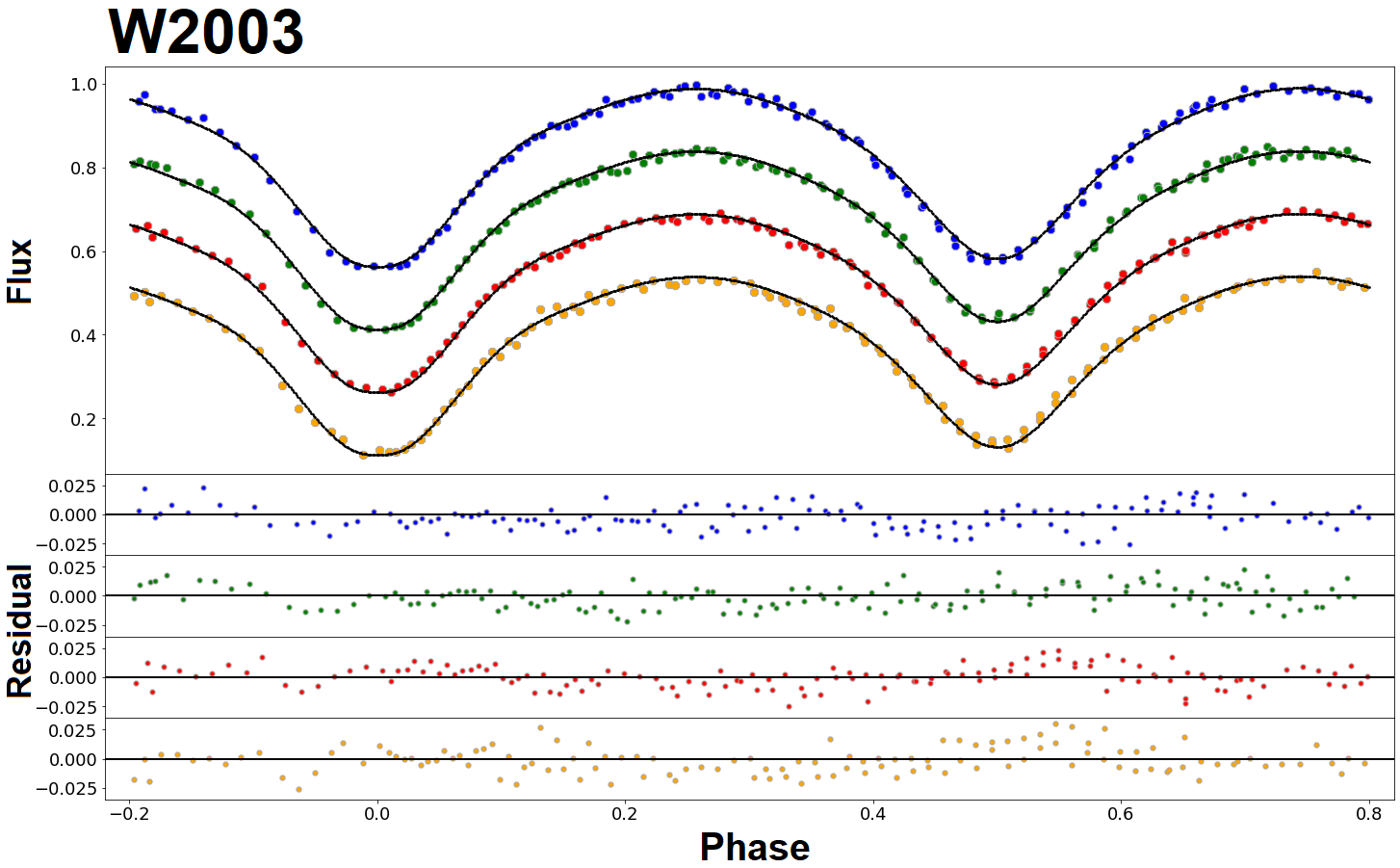}
\includegraphics[width=0.495\textwidth]{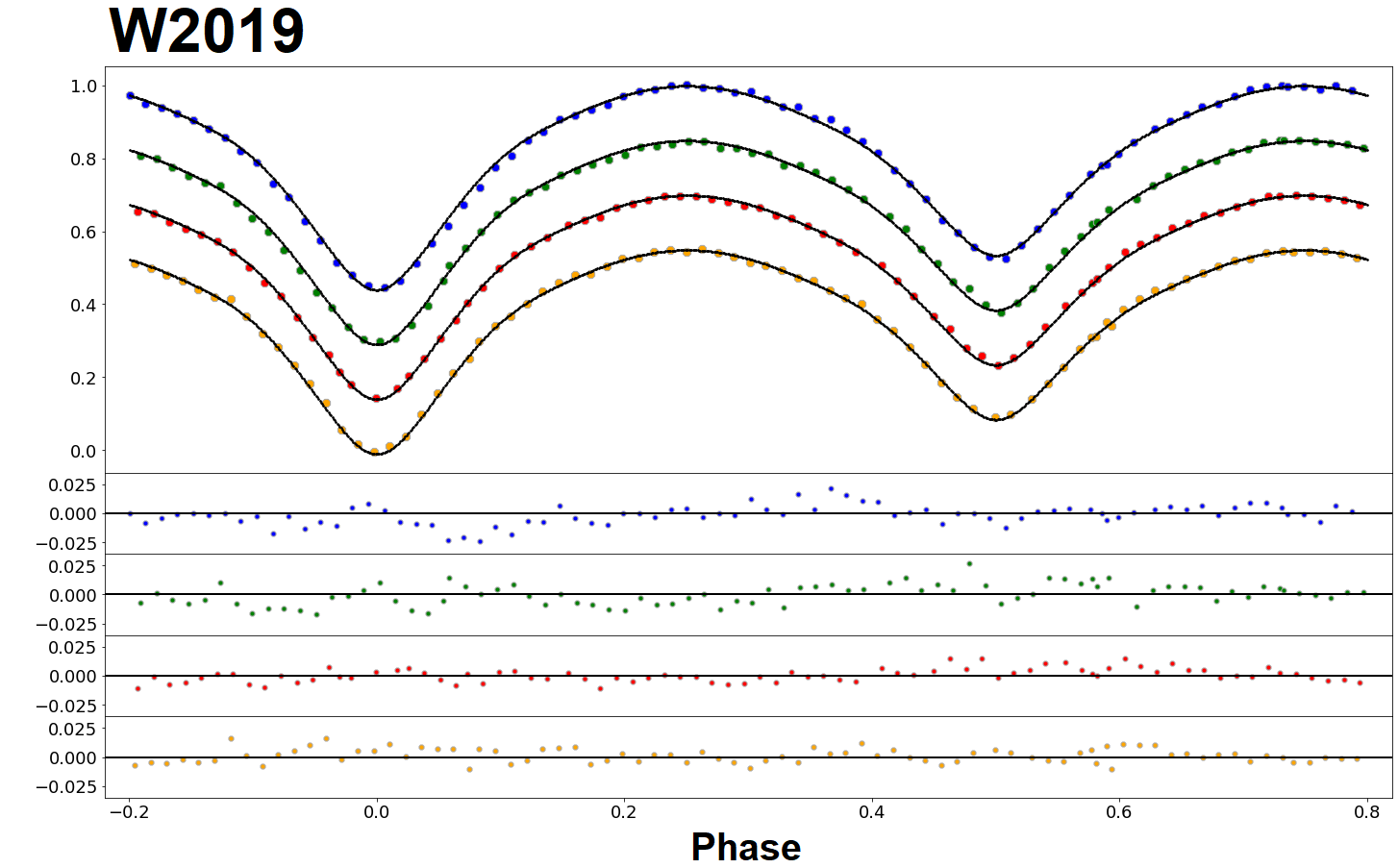}
\caption{The colored dots represent the observed light curves of the systems in different filters, and the synthetic light curves, generated using the light curve solutions, are also shown.}
\label{LCs}
\end{figure*}

\begin{figure*}
\centering
\includegraphics[width=0.89\textwidth]{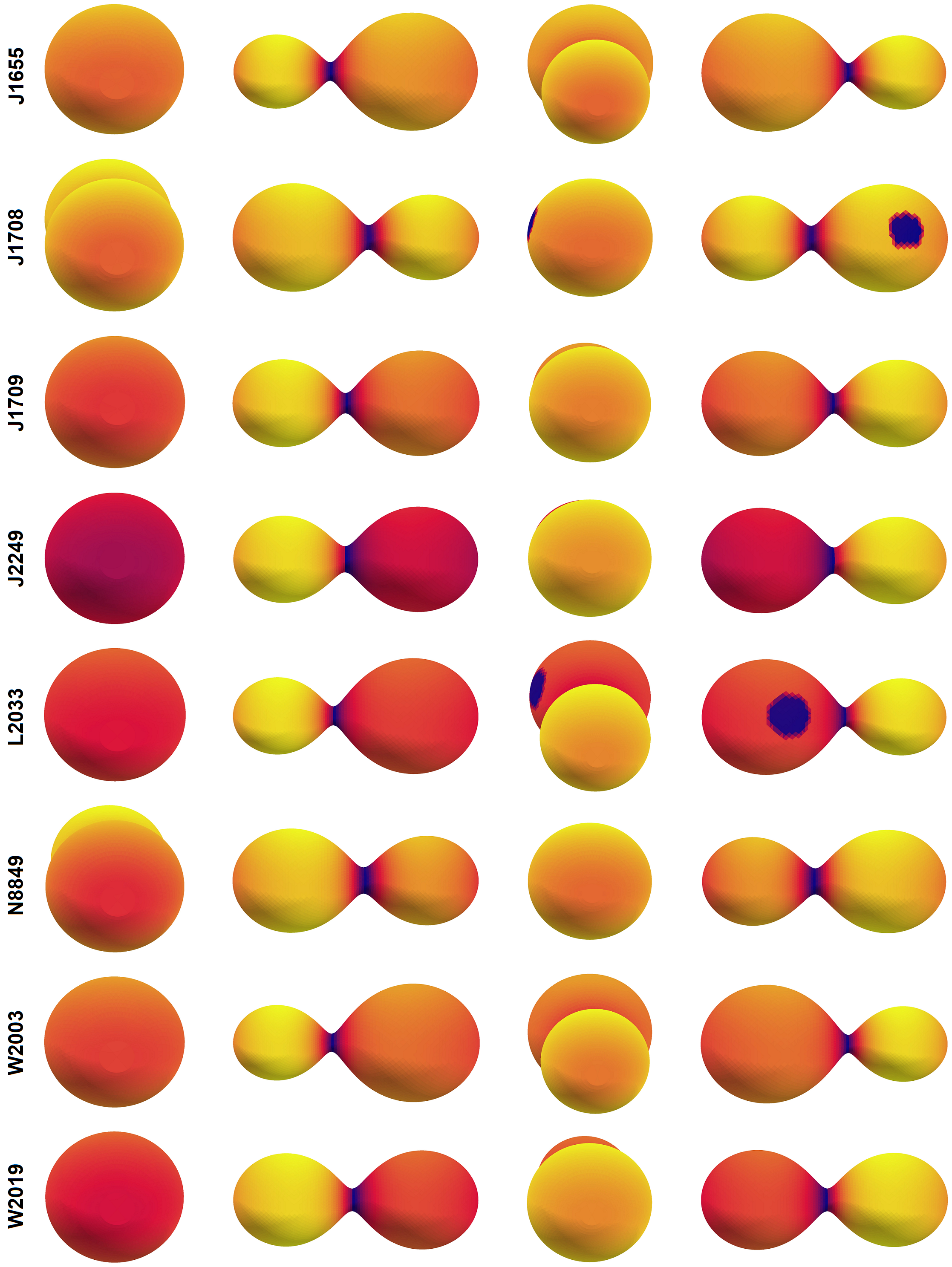}
\caption{Three-dimensional views of the stars in the target binary systems at orbital phases 0, 0.25, 0.5 and 0.75, respectively. The colors represent variations in surface temperature based on model simulations, with darker tones indicating cooler regions and lighter areas corresponding to higher temperatures.}
\label{3d}
\end{figure*}

Table \ref{lc-analysis} contains the final results of the light curve solutions. The parameters used to characterize a starspot, such as colatitude (Col.$^\circ$), longitude (Long.$^\circ$), angular radius (Radius$^\circ$), and the temperature ratio ($T_{spot}/T_{star}$), are presented in Table \ref{lc-analysis}. Figure \ref{LCs} displays the observed and synthetic light curves of the binary systems. Figure \ref{3d} illustrates the three-dimensional representations of the binary systems.

\begin{table*}
\renewcommand\arraystretch{1.8}
\caption{Light curve solutions of the target binary stars.}
\centering
\begin{center}
\footnotesize
\begin{tabular}{c c c c c c c c c}
\hline
Parameter & J1655 & J1708 & J1709 & J2249 & L2033 & N8849 & W2003 & W2019\\
\hline
$T_{1}$ (K) & 4823(35) & 5076(29) &	5272(35) & 5228(26) & 4790(37) & 5282(37) & 5699(26) & 5411(70)\\
$T_{2}$ (K) & 4761(29) & 5075(28) &	5094(29) & 4729(28) & 4493(51) & 5188(32) & 5514(30) & 5110(52)\\
$q=M_2/M_1$ & 2.72(1) & 0.60(2) & 1.48(4) & 1.51(9) & 2.39(7) &	0.72(6) &	2.63(8) & 1.14(8)\\
$i^{\circ}$ & 80.6(3) & 82.2(6) & 85.9(7) & 87.4(8) & 76.4(9) &	81.6(6) &	80.3(4) & 82.0(7)\\
$f$ & 0.11(2) & 0.15(1) & 0.10(1) & 0.17(1) & 0.11(2) &	0.16(2) & 0.10(2) & 0.11(2)\\
$\Omega_1=\Omega_2$ & 6.2(2) & 3.0(2) & 4.4(2) & 4.4(2) & 5.7(1) & 3.2(1) & 6.1(2) & 3.9(2)\\
$l_1/l_{tot}$($V$) & 0.31(1) &	0.61(1) & 0.46(1) & 0.56(1) &	0.41(1) &  0.60(1) & 0.33(1) & 0.55(1)\\
$l_2/l_{tot}$($V$) & 0.69(1) &	0.39(1) & 0.54(1) & 0.44(1) &	0.59(1) &	0.40(1) & 0.67(1) & 0.45(1)\\
$r_{(mean)1}$ &	0.30(1) & 0.44(1) & 0.35(1) & 0.36(1) & 0.31(1) & 0.42(1) & 0.31(1) & 0.38(1)\\
$r_{(mean)2}$ &	0.48(1) & 0.35(1) & 0.42(1) & 0.43(1) & 0.47(1) & 0.37(1) & 0.47(1) & 0.40(1)\\
\hline
$Col.^\circ$(spot) & - & 75(2) & - & - & 104(3) & - & - & -\\
$Long.^\circ$(spot) & - & 102(3) &	- &	- &	293(3) & - & - & -\\
$Radius^\circ$(spot) & - & 15(1) &	- &	- &	18(1) & - & - & -\\
$T_{spot}/T_{star}$ & - & 0.88(1) & - & - & 0.88(1) &	- &	- & -\\
Component &	- &	Primary & - & - & Secondary &	- &	- & -\\
\hline
\end{tabular}
\end{center}
\label{lc-analysis}
\end{table*}

\vspace{0.6cm}
\section{Absolute Parameters Estimations}
\label{sec6}
We employed Gaia DR3 parallaxes to derive the absolute parameters of the targets binary stars, which makes this approach particularly effective when only photometric observations are available \citep{2024NewA..11002227P}. To verify the appropriateness of using Gaia DR3 parallaxes for our systems, we initially calculated and assessed the interstellar extinction ($A_V$). Based on the findings of \cite{2024PASP..136b4201P}, $A_V$ values exceeding approximately 0.4 are considered unsuitable. Using the 3D dust maps provided by \citet{2019ApJ...887...93G}, we determined the extinction values and found that six systems met this criterion (Table \ref{absolute}). However, W2003 and W2019, with $A_V$ values of 0.824(10) and 2.002(18), respectively, were not considered for absolute parameter estimation using this method, as their high extinction values exceed the reliability threshold for the Gaia DR3 parallax-based approach.

The absolute magnitude of each system ($M_V$) was estimated using its observed maximum brightness $V_{\text{max}}$, the Gaia DR3 distance, and the corresponding $A_V$ value. The $V_{\text{max}}$ values used in this calculation were taken from our observational data (Table \ref{absolute}). Next, we calculated the individual absolute magnitudes $M_{V1}$ and $M_{V2}$ using the luminosity ratio $l_{1,2}/l_{\text{tot}}$ derived from the $V$-band light curve solutions. The absolute bolometric magnitudes ($M_{\text{bol},1}$ and $M_{\text{bol},2}$) were then obtained using bolometric corrections ($BC_1$ and $BC_2$) following the calibration of \citet{1996ApJ...469..355F}. The stellar luminosities were derived based on the relationship between absolute bolometric magnitude and luminosity, adopting a solar bolometric magnitude of $M_{\text{bol}\odot} = 4.73$ mag \citep{2010AJ....140.1158T}. Furthermore, with known luminosities and effective temperatures (from the light curve solutions), stellar radii ($R$) were computed.

The semi-major axis $a$ of each binary system was estimated by combining the mean fractional radii ($r_{\text{mean},1,2}$) with the computed radii ($R_{1,2}$), and then averaging $a_1$ and $a_2$. Finally, using the derived values of $a$, orbital period $P$, and mass ratio $q$, the individual stellar masses were calculated by using Kepler's third law:

\begin{eqnarray}
M{_1}=\frac{4\pi^2a^3}{GP^2(1+q)}\label{eq:M1},\\
M{_2}=q\times{M{_1}}\label{eq:M2}.
\end{eqnarray}

The surface gravity ($g$) of the stars was estimated on a logarithmic scale, based on the calculated mass and radius values. Additionally, the orbital angular momentum ($J_0$) was calculated based on the total mass, mass ratio, and orbital period of the systems, using Equation \ref{eqJ0} from the \cite{2006MNRAS.373.1483E} study.

\begin{equation}\label{eqJ0}
J_0=\frac{q}{(1+q)^2} \sqrt[3] {\frac{G^2}{2\pi}M^5P}
\end{equation}

Table \ref{absolute} presents the resulting absolute parameters for the six contact binary systems.

\begin{table*}
\renewcommand\arraystretch{1.5}
\caption{Estimated absolute parameters of the systems.}
\centering
\begin{center}
\footnotesize
\begin{tabular}{c c c c c c c}
\hline
Parameter & J1655 & J1708 & J1709 & J2249 & L2033 & N8849\\
\hline
$M_1(M_\odot)$&0.32(8)&0.89(19)&0.59(19)&0.63(8)&0.48(13)&0.92(6)\\
$M_2(M_\odot)$&0.9(2)&0.53(12)&0.9(3)&0.96(12)&1.1(3)&0.66(4)\\
$R_1(R_\odot)$&0.57(6)&0.89(9)&0.72(9)&0.81(6)&0.66(6)&0.90(5)\\
$R_2(R_\odot)$&0.89(9)&0.71(7)&0.86(11)&0.98(8)&0.98(11)&0.78(4)\\
$L_1(L_\odot)$&0.16(3)&0.47(9)&0.36(9)&0.44(6)&0.21(3)&0.56(5)\\
$L_2(L_\odot)$&0.37(7)&0.30(5)&0.45(11)&0.43(6)&0.35(7)&0.39(3)\\
$M_{bol1}(mag)$&6.76(19)&5.55(19)&5.9(3)&5.63(14)&6.45(18)&5.36(10)\\
$M_{bol2}(mag)$&5.83(19)&6.05(19)&5.62(26)&5.65(15)&5.88(19)&5.76(9)\\
$log(g)_1(cgs)$&4.44(19)&4.49(19)&4.5(3)&4.42(12)&4.48(19)&4.49(8)\\
$log(g)_2(cgs)$&4.48(19)&4.46(19)&4.5(3)&4.44(12)&4.51(21)&4.48(8)\\
$a(R_\odot)$&1.87(16)&2.04(15)&2.0(2)&2.27(10)&2.11(18)&2.13(5)\\
$logJ_0$&51.33(19)&51.53(17)&51.6(2)&51.6(9)&51.58(19)&51.63(5)\\
\hline
$A_V(mag)$ &	0.105(1) &	0.236(1) &	0.268(1) &	0.199(1) &	0.115(1) &	0.266(1)\\
$V_{max}(mag.)$ & 15.83(13) & 14.99(16) & 16.27(14) & 15.01(10) & 16.38(10) & 13.26(9)\\
$BC_1(mag.)$&-0.395(19)&-0.275(12)&-0.203(11)&-0.217(9)&-0.413(19)&-0.199(12)\\
$BC_2(mag.)$&-0.429(17)&-0.276(11)&-0.268(12)&-0.448(17)&-0.61(4)&-0.232(12)\\
\hline
\end{tabular}
\end{center}
\label{absolute}
\end{table*}

\vspace{0.6cm}
\section{Discussion and Conclusion}
\label{sec7}
This study provides the first light curve analysis, investigation of orbital period variations, and estimation of absolute parameters for eight contact binary systems. The following discussions and conclusions are derived from the analysis of our results:

A) We have investigated the orbital period variations in eight target systems. The main causes of these variations include apsidal motion, magnetic activity, the third-body effect, and the transfer or loss of mass and angular momentum (\citealt{2024NewA..10502112S}).

All the eight targets shows long-term variations in their orbital periods. We assumed that the long-term variations are due to mass transfer between the two components of the binary systems. We can use the following Equation \citep{k1958} to calculate the mass transfer rate,

\begin{equation}
\begin{aligned}
\frac{\dot{P}}{P}=-3\dot{M}(\frac{1}{M_1}-\frac{1}{M_2}).
\end{aligned}
\end{equation}

Given that the absolute parameters of the six target systems were estimated in Section 6, their mass transfer rates were subsequently calculated and are presented in Table \ref{tabO-C}.

B) For the J1708 and L2033 systems, the light curve analyses required the inclusion of a cold starspot on one of the components to account for the observed asymmetry between the maxima. This asymmetry is a characteristic feature of the O'Connell effect, a recognized phenomenon that modifies the shape and symmetry of light curves in contact binary systems (\citealt{1951PRCO....2...85O}). The stellar temperatures of the analyzed systems range from 4493 K to 5699 K, as derived from our results. The temperature differences ($\Delta T = |T_1 - T_2|$) for each system, as derived from the light curve analysis in this study, are presented in Table \ref{conclusion}. The uncertainties of the temperature differences were calculated by adding the individual temperature errors in quadrature. The spectral classifications of the component stars were determined using the \cite{2000asqu.book.....C} and \cite{2018MNRAS.479.5491E} studies (Table \ref{conclusion}).

C) The fillout factor is a parameter used to describe the degree of contact in close binary systems, where both stars share a common envelope. It indicates how much the stars exceed their Roche lobe boundaries and thus how strongly they are interacting. A low fillout factor suggests a shallow contact configuration, while higher values point to more significant overcontact. This factor helps classify the evolutionary state of the system and provides insights into energy transfer and mass exchange processes between the stellar components.

Contact binary systems are categorized according to their fillout factor into three classes: deep ($f \geq 50\%$), medium ($25\% \leq f < 50\%$), and shallow ($f < 25\%$) systems \citep{2022AJ....164..202L}. Therefore, based on the light curve solutions, all target systems fall into the shallow category.

\begin{figure*}
\centering
\includegraphics[width=0.49\textwidth]{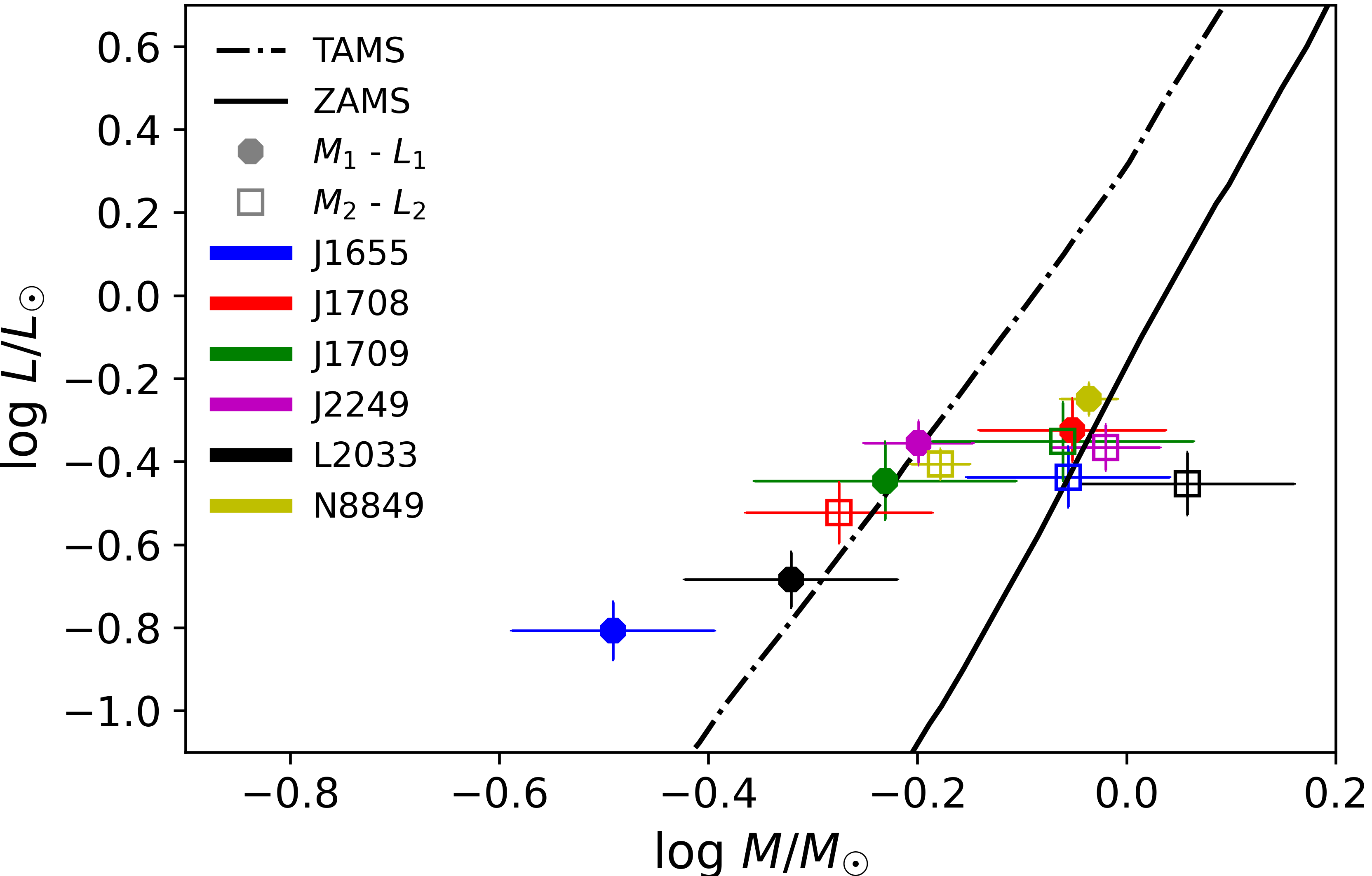}
\includegraphics[width=0.49\textwidth]{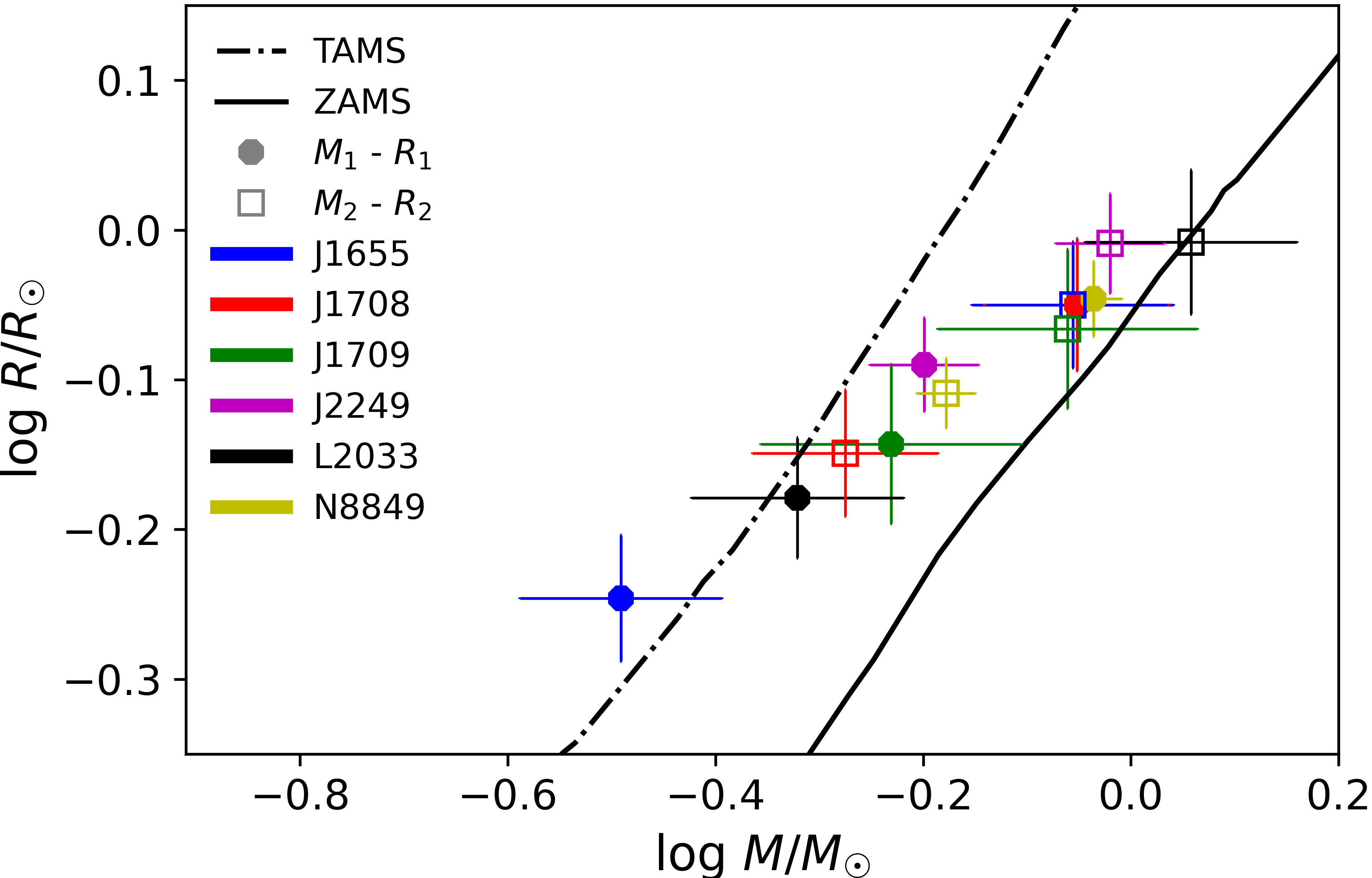}
\caption{Position of both stars in six target system on the $M$–$R$ and $M$–$L$ diagrams.}
\label{MLR}
\end{figure*}

D) The evolutionary status of the target systems were illustrated using logarithmic Mass–Radius ($M$–$R$) and Mass–Luminosity ($M$–$L$) diagrams, constructed based on the derived absolute parameters for six of targets (Table \ref{absolute}, Figure \ref{MLR}). In these diagrams, the stellar components are shown relative to the Zero-Age Main Sequence (ZAMS) and Terminal-Age Main Sequence (TAMS) lines, as presented by the \cite{2000AAS..141..371G} study.

Based on the light curve analysis and the estimated absolute parameters, four of the systems contain stars in which the less massive component has a higher temperature (W-subtype). In contrast, in two systems, the hotter component is also the more massive star (A-subtype). In the absence of mass estimates for the W2003 and W2019 systems, their subtypes could not be classified at this stage. The subtype determine of six system are presented in Table \ref{conclusion}. As illustrated in Figure \ref{MLR}, the lower-mass components tend to be located near the TAMS, while the more massive stars are situated closer to the ZAMS.

A few components are located slightly above the TAMS or below the ZAMS in Figure \ref{MLR}. However, when considering the uncertainties in the derived parameters, all of them remain consistent with the main-sequence region within their error margins. Stars appearing above the TAMS may reflect evolutionary processes such as mass transfer, which can increase their luminosities or temperatures beyond the expectations from single-star tracks. Similarly, stars below the ZAMS might result from mass transfer processes, which can increase a star's mass while its structure and temperature still reflect a less massive evolutionary state. These trends have been identified in studies of contact binaries, reflecting their complex nature and the limits of single-star evolutionary models in accurately describing the structure and evolution of contact systems (e.g. \citealt{2005ApJ...629.1055Y}, \citealt{2006AcA....56..199S}, \citealt{2008MNRAS.387...97L}).

E) Investigating the evolution of W UMa-type binary systems involves identifying the pathways through which stars come to fill their Roche lobes. These evolutionary routes are predominantly shaped by nuclear and angular momentum evolution, both of which are closely tied to the stellar mass \citep{1989MNRAS.237..447H}. As such, the initial masses of the binary components are essential parameters for understanding the formation history and subsequent evolution of these systems.

In this work, the initial masses of the primary ($M_{1i}$) and secondary ($M_{2i}$) stars for six targets were determined, following the approach outlined by \cite{2013MNRAS.430.2029Y}.

To begin with, the initial mass of the secondary component was derived using Equation \ref{Mi2}, as presented in \cite{2013MNRAS.430.2029Y}:

\begin{equation}\label{Mi2}
M_{2i}=M_2+\Delta M=M_2+2.50(M_L-M_2-0.07)^{0.64}
\end{equation}

\noindent where $M_2$ denotes the current mass of the secondary star, $\Delta M$ represents the amount of mass transferred to the secondary component (i.e., the increase in mass compared to its initial value), while $M_L$ is obtained from the mass-luminosity relation shown in Equation \ref{M_L}:

\begin{equation}\label{M_L}
M_L=\left(\frac{L_2}{1.49}\right)^{\frac{1}{4.216}}
\end{equation}

Subsequently, the initial mass of the primary star was computed using Equation \ref{Mi1}:

\begin{equation}\label{Mi1}
M_{1i}=M_1-(\Delta M-M_{\text{lost}})=M_1-\Delta M(1-\gamma)
\end{equation}

All quantities in Equations \ref{Mi2}, \ref{M_L}, and \ref{Mi1} are expressed in solar units. The parameter $M_{\text{lost}}$ represents the total mass lost of each system, and $\gamma$ is the ratio of $M_{\text{lost}}$ to $\Delta M$:

\begin{equation}\label{Mlost}
M_{\text{lost}}=\gamma \times \Delta M
\end{equation}

In line with the findings of \cite{2013MNRAS.430.2029Y}, we adopted $\gamma = 0.664$. The reciprocal of the initial mass ratio ($1/{q_i}$) was utilized as a constraint in estimating the initial mass of the primary component. The estimated values for initial masses and mass loss across the six systems are provided in Table \ref{conclusion}.

As presented in Table \ref{conclusion}, the initial masses of the primary components in the analyzed systems fall within the range of 0.6–1 $M_{\odot}$. Systems with $M_{1i}$ values between 0.2 and 1.5 $M_{\odot}$ are likely to have undergone relatively rapid angular momentum loss, consistent with the scenario proposed by \cite{2004ARep...48..219T}. Additionally, our estimations indicate that the initial masses of the secondary components lie between 0.6 and 1.2 $M_{\odot}$. Notably, these secondary stars differ structurally from standard main-sequence single-stars, as their initial masses exceed their current values \citep{2013MNRAS.430.2029Y}. The estimated mass loss for the target systems aligns well with the findings of \cite{2013MNRAS.430.2029Y}.

\begin{table*}
\renewcommand\arraystretch{1.5}
\caption{Some conclusions regarding the target systems.}
\centering
\begin{center}
\footnotesize
\begin{tabular}{c c c c c c c c c}
\hline
Parameter & J1655 & J1708 & J1709 & J2249 & L2033 & N8849 & W2003 & W2019\\
\hline
$\Delta T=|T_1-T_2|$ ($K$) & 62(45) & 1(40) & 178(45) & 499(38) & 297(63) & 94(49) & 185(40) & 301(87)\\
Spectral category & K2-K3 & K1-K1 & K0-K1 & K0-K3 & K2-K5 & K0-K0 & G6-G8 & G8-K1\\
Subtype & W & A & W & W & W & A & - & -\\

$M_{1i}$ ($M_{\odot}$) & 0.6(2) & 0.7(3) & 0.7(4) & 0.8(2) & 1.0(3) & 0.9(1) & - & -\\

$M_{2i}$ ($M_{\odot}$) & 1.2(2) & 1.0(4) & 1.0(6) & 1.0(3) & 1.0(4) & 0.6(1) & - & -\\

$M_{\text{lost}}$ ($M_{\odot}$) & 0.6(2) & 0.3(3) & 0.3(5) & 0.2(2) & 0.3(3) & 0.1(1) & - & -\\

\hline
\end{tabular}
\end{center}
\label{conclusion}
\end{table*}

\vspace{0.6cm}
\section*{Data Availability}
Ground-based data are available in the paper's online supplement.

\vspace{0.6cm}
\section*{Acknowledgments}
This manuscript, including the observation, analysis, and writing processes, was provided by the BSN project (\url{https://bsnp.info/}). Work by Kai Li was supported by the National Natural Science Foundation of China (NSFC) (No. 12273018) and by the Qilu Young Researcher Project of Shandong University. This paper is based on observations carried out at the Observatorio Astron\'omico Nacional on the Sierra San Pedro M\'artir which is operated by the Universidad Nacional Aut\'onoma de M\'exico. We used IRAF, distributed by the National Optical Observatories and operated by the Association of Universities for Research in Astronomy, Inc., under a cooperative agreement with the National Science Foundation. We used data from the European Space Agency mission Gaia (\url{http://www.cosmos.esa.int/gaia}). The authors would like to express their gratitude to Dr. David Valls-Gabaud for all his help and advice.

\vspace{0.6cm}
\bibliography{REFS}{}
\bibliographystyle{aasjournal}

\end{document}